\documentclass[pre,twocolumn]{revtex4}

\usepackage{graphicx}
\usepackage{amsopn}
\usepackage{amsfonts}
\usepackage{amsmath}
\usepackage{latexsym}
\usepackage{amsmath,bm,amssymb}
\usepackage{pgfplots}
\usepackage{color}

\begin{document}

\title{Correlations in a weakly interacting two-dimensional random flow}

\author{I.V. Kolokolov$^{1,2}$, V.V. Lebedev$^{1,2}$, V.M. Parfenyev$^{1,2}$}

\affiliation{$^1$Landau Institute for Theoretical Physics, RAS, \\
142432, Chernogolovka, Moscow region, Russia; \\
$^2$National Research  University Higher School of Economics, \\
101000, Myasnitskaya ul. 20, Moscow, Russia.}

\date{\today}

\begin{abstract}

We analytically examine fluctuations of vorticity excited by an external random force in two-dimensional fluid. We develop the perturbation theory enabling one to calculate nonlinear corrections to correlation functions of the flow fluctuations found in the linear approximation. We calculate the correction to the pair correlation function and the triple correlation function. It enables us to establish the criterion of validity of the perturbation theory for different ratios of viscosity and bottom friction. We find that the corrections to the second moment are anomalously weak in the cases of small bottom friction and small viscosity and relate the weakness to the energy and enstrophy balances. We demonstrate that at small bottom friction the triple correlation function is characterized by universal scaling behavior in some region of lengths. The developed perturbation method was verified and confirmed by direct numerical simulations.

\end{abstract}

\maketitle

\section{Introduction}
\label{sec:intro}

For a three-dimensional flow in a fluid, the only dissipation mechanism of the kinetic energy is viscosity. Thus, there is the only dimensionless parameter, Reynolds number, characterizing the degree of nonlinearity. In contrast, when considering effectively two-dimensional thin fluid films one has to deal with two dissipative mechanisms: viscosity and bottom friction. An interplay of the dissipative mechanisms leads to a more complicated characterization of the flow nonlinearity.

The traditional way to produce flow in $2d$ hydrodynamics is to apply an external periodic static force (Kolmogorov force) to the fluid. Such setup is used both in laboratory experiments with thin liquid films \cite{Sommeria86,Tabeling02,Shats09,Orlov17} and in numerical simulations \cite{MCH04,Mishra15,DFKL22}. Then the transition to turbulence appears to be complicated, it can be soft transition or jump (depending on the ratio between the pumping length and the box size) and can go through some intermediate stages. Aiming to simplify the situation and to reveal its universal features, here we analytically consider the case when the external force exciting the flow is a random function of space and time with homogeneous statistical properties. Our preliminary and yet unpublished results show that the transition to turbulence in this case occurs gradually (without jumps). Therefore, there is a wide range of parameters where the interaction of the random flow fluctuations is relatively weak.

The regime of relatively weak random flow can be easily realized numerically, and we use direct numerical simulations to verify our analytical findings. Besides, the regime can be realized experimentally in thin fluid films by applying a random rapidly varying in time external force to the fluid. Unfortunately, we do not know works devoted to such setup. An interest to this regime in thin fluid films may be related to the organization of a controllable mixing process of impurities in the film. To characterize the process one has to find the coefficient of turbulent diffusion.

In this paper, we investigate effects related to correlations in a weakly interacting $2d$ random flow, excited by an external force, random in space and time. The character of the flow interaction essentially depends on the ratio of the bottom friction coefficient $\alpha$ and $\nu k_f^2$, where $\nu$ is the kinematic viscosity coefficient and $k_f$ is the characteristic wave vector of the external force. We examine an arbitrary ratio between $\alpha$ and $\nu k_f^2$ in our work. Of course, arbitrary values of the coefficients $\nu$ and $\alpha$ can be easily implemented in numerical simulations of the two-dimensional fluid. As for experiments, the situation is as follows.

For thin liquid films it is easy to achieve the inequality $\alpha\gg \nu k_f^2$, since $\alpha$ can be estimated as $\nu/ h^2$, where $h$ is the thickness of the film. The characteristic length of the pumping force should be larger than $h$ (otherwise the flow cannot be treated as two-dimensional), thus $k_f h \lesssim 1$. That leads to the conclusion $\alpha \gtrsim \nu k_f^2$. However, to observe interesting effects related, e.g., to large-scale vortices in the turbulent regime, one should make $\alpha$ as small as possible. To achieve the goal some experimental tricks are used related to multilayered films \cite{Tabeling02,Shats09,Orlov17}. The opposite inequality $\alpha\ll \nu k_f^2$ can be achieved for the effectively two-dimensional subsystem of the fast rotating three-dimensional fluid where the effective bottom friction is related to formation of Ekman boundary layers~\cite{Vergeles21}. Other cases, where small $\alpha$ can be realized, include suspended soap or smectic films, which have no contact with a solid boundary and therefore do not exhibit a velocity gradient across the film thickness \cite{kellay2002two, Yablonskii2017}.

Our aim is to examine consistently first corrections to correlation functions of the flow fluctuations. More precisely, we analyze corrections to the pair correlation function of vorticity and its triple correlation function. It enables one to establish the criterion of validity of the perturbation theory, and, consequently, to predict conditions of exciting true turbulent state. Note also, that the flow fluctuations are believed to be weakly interacting inside the coherent vortices appearing in $2d$ turbulence \cite{KL15,kolokolov2016structure,kolokolov2016velocity,frishman17}. This work will help to clarify validity of the approach.

The rest of the paper is organized as follows. In Sec.~\ref{sec:general}, we give the general theoretical framework for the two-dimensional flow excited by an external force and discuss the linear approximation. Sec.~\ref{sec:perturb} is devoted to the definition of correlation functions using functional integration. In Sec.~\ref{sec:firstcorr}, we discuss the development of perturbation series enabling one to calculate effects related to the nonlinear interaction and calculate first corrections to the correlation functions. In particular, in Sec.~\ref{sec:paircorr} we analyze the correction to the pair correlation function of vorticity, thereby establishing parameters controlling the perturbation series, and in Sec.~\ref{sec:secondmom} we examine the correction to the second moment of vorticity, which has some features. Sec.~\ref{sec:triplecorr} is devoted to the calculation of the triple correlation function of vorticity. In Sec.~\ref{sec:directns}, we briefly expose results of direct numerical simulations confirming our analytical analysis. In Conclusion, we outline the results of our work and discuss some perspectives.

\section{General relations}
\label{sec:general}

In this study, we consider an unbounded two-dimensional fluid. We are interested in fluctuations of the flow excited by an external, relatively weak, random force, and examine effects related to their nonlinear interaction. We take into account two dissipation mechanisms: bottom friction and viscosity. The external force is assumed to be a random function of time and coordinates possessing statistical properties homogeneous in time and space.

The flow excited by the external force is described by a (two-dimensional) velocity field $\bm v$, which is a random function of time and coordinates, as well as of the external force. The flow is assumed to be incompressible, $\nabla \cdot \bm v=0$. We examine the statistically stationary state of the fluid. It is characterized by correlation functions of the velocity. For example, one can be interested in the pair correlation function $\langle \bm v(\bm r) \cdot \bm v(\bm x) \rangle$ for the velocities taken at the points $\bm r$ and $\bm x$. The angular brackets here and below indicate averaging over the system statistics.

Having in mind dynamics of thin films, one should add to the two-dimensional Navier-Stokes equation the term related to bottom friction:
\begin{equation}
\partial_t \bm v +(\bm v \cdot \nabla) \bm v +\nabla p
=-\alpha \bm v +\nu \nabla^2 \bm v +\bm f.
\label{basic0}
\end{equation}
Here $p$ is pressure, $\alpha$ is the bottom friction coefficient and $\nu$ is the kinematic viscosity coefficient. Taking the divergence from equation (\ref{basic0}), we arrive at the following relation for pressure $p$:
\begin{equation}
(\partial_\alpha v_\beta)(\partial_\beta v_\alpha) +\nabla^2 p=\nabla \bm f,
\end{equation}
where Greek indices run over $\{1,2\}$, and we sum over the repeated indices.

In two dimensions, it is convenient to describe the flow in terms of vorticity $\varpi$:
\begin{equation}
\varpi= \mathrm{curl}\, \bm v
\equiv \partial_1 v_2-\partial_2 v_1.
\label{varpi}
\end{equation}
Obviously, $\varpi$ is a scalar (or, more precisely, a pseudoscalar) field. The equation controlling the flow is derived from Eq. (\ref{basic0}) by taking curl of it:
\begin{equation}
\partial_t\varpi+ \bm v \cdot \nabla \varpi
= -\alpha \varpi +\nu \nabla^2 \varpi
+\mathrm{curl}\, \bm f .
\label{basic}
\end{equation}
Note that the pressure term drops from the equation.

To close equation (\ref{basic}), one should restore the velocity field $\bm v$ from the vorticity field $\varpi$. Due to the incompressibility condition $\partial_1 v_1+\partial_2 v_2=0$, it is possible to introduce the stream function $\Psi$, related to the velocity components and to the vorticity as
\begin{equation}
v_1=\frac{\partial \Psi}{\partial x_2}, \quad
v_2=-\frac{\partial \Psi}{\partial x_1}, \quad
\varpi=-\nabla^2 \Psi.
\label{stre1}
\end{equation}
To determine the stream function, it is necessary to solve the Laplace equation $\nabla^2 \Psi=-\varpi$. Its solution can be written, say, as the integral
\begin{equation}
\Psi(\bm r)=-\frac{1}{2\pi}
\int d^2 x\, \varpi(\bm x) \ln |\bm r-\bm x|.
\label{stre2}
\end{equation}
After calculating the integral one finds the velocity components in accordance with Eq. (\ref{stre1}), thus expressing the velocity via the vorticity.

Multiplying equation (\ref{basic0}) by $\bm v$ and averaging it, one obtains the energy balance
\begin{equation}
\langle \bm f \cdot \bm v \rangle= \alpha \langle v^2 \rangle +\nu \langle \varpi^2 \rangle.
\label{energyb}
\end{equation}
We introduce the special notation for the average in the left hand side of Eq.~(\ref{energyb}), $\epsilon =\langle \bm f \cdot \bm v \rangle$. It is the average power of the external force per unit mass. At deriving Eq.~(\ref{energyb}) we omitted all full derivatives over time and coordinates, having in mind homogeneity in space and time. The relation (\ref{energyb}) has simple physical meaning: the energy pumped into the fluid is dissipated by bottom friction and viscosity.

Analogously, multiplying equation (\ref{basic}) by $\varpi$ and averaging, one obtains
\begin{equation}
\langle \phi \varpi \rangle=
\alpha \langle \varpi^2 \rangle
+\nu \langle (\nabla \varpi)^2 \rangle,
\label{enstrophyb}
\end{equation}
where $\phi=\mathrm{curl}\, \bm f$. The physical meaning of the relation (\ref{enstrophyb}) is the enstrophy balance. The enstrophy production rate $\langle \phi \varpi \rangle$ can be estimated as $\epsilon k_f^2$ where $k_f$ is the characteristic wave vector of the pumping. The relations (\ref{energyb}) and (\ref{enstrophyb}) are exact and valid for any pumping level.

In the paper, we exploit a model in which the external force $\bm f$ is shortly correlated in time and has zero mean, $\langle \bm f\rangle=0$. Then statistical properties of pumping are determined exclusively by the pair correlation function of the force
\begin{equation}
\langle f_\alpha (t, \bm x) f_\beta (\tau, \bm y) \rangle
= 2\epsilon \delta_{\alpha\beta} \delta(t-\tau) \varXi(\bm x-\bm y).
\label{force1}
\end{equation}
This expression implies homogeneity in space and time. Here $\varXi(\bm r)$ is a function of coordinates determined by details of pumping. To ensure the property $\epsilon= \langle \bm f \cdot \bm v \rangle$, one should take $\varXi(\bm 0)=1$. Moving on to the quantity $\phi= \mathrm{curl}\, \bm f$, which appears in equation (\ref{basic}) for vorticity, one finds from expression~(\ref{force1})
\begin{equation}
\langle\phi(t,\bm x)\phi(\tau,\bm y)\rangle
=-2\epsilon\delta(t-\tau) \nabla^2 \varXi(\bm x-\bm y),
\label{basic2}
\end{equation}
where the gradient $\nabla$ can be taken either over $\bm x$ or over $\bm y$. Note that the enstrophy production rate per unit mass $\langle \phi \varpi \rangle$ in our model is unambiguously determined by the function $\varXi$,
\begin{equation}
\langle \phi \varpi \rangle=-\epsilon \nabla^2 \varXi(\bm 0),
\label{enstrophy}
\end{equation}
as it follows from Eqs.~(\ref{force1}) and (\ref{basic2}).

We assume that the pumping is statistically isotropic, i.e. $\varXi(\bm r)$ depends only on the absolute value of $\bm r$. Next, we assume that the function $\varXi(\bm r)$ has a characteristic length $k_f^{-1}$ (correlation length of the pumping), where $k_f$ is the characteristic wave vector of the force $\bm f$. As an example, one can consider the expression
\begin{equation}
\varXi(\bm r)=\exp\left(- k_f^2 r^2/2\right),
\label{farce1}
\end{equation}
which satisfies the condition $\varXi(\bm 0)=1$ and decays fast at large $r$. Its Fourier transform is given by
\begin{equation}
\tilde \varXi(\bm k)=\frac{2\pi}{k_f^2}
\exp\left(-\frac{k^2}{2 k_f^2}\right),
\label{gausspu}
\end{equation}
and it decays fast as the wave vector $k$ grows. It is also instructive to consider pumping, which has a narrow spectrum $\tilde \varXi(\bm k)$ in the Fourier representation. As we will see later, some qualitative differences may arise in this case. To quantitatively illustrate this scenario, we consider the infinitely thin function
\begin{equation}
\tilde\varXi(\bm k) = 2 \pi \epsilon \delta(k-k_f)/k_f,
\label{narrow}
\end{equation}
satisfying the condition $\varXi(\bm 0)=1$ in the real space.

Finally, our approach implies that the system size $L$ is much larger than all characteristic length scales of the problem. First of all, $L$ should be much larger than the pumping length $k_f^{-1}$, $k_f L\gg 1$. However, as we will show below, some quantities can be determined by the scales of the order of $\sqrt{\nu/\alpha}$, which may be larger than the pumping length $k_f^{-1}$. For validity of our results, these scales should be much smaller than $L$ as well.

\subsection{Linear approximation}\label{sec:linear}

In the linear approximation, the equation (\ref{basic}) is reduced to
\begin{equation}
(\partial_t+\alpha -\nu \nabla^2)\varpi=\phi.
\label{basic1}
\end{equation}
To find correlation functions of $\varpi$ in this approximation, one should solve equation~(\ref{basic1}) for arbitrary $\phi$ and then average the corresponding product over the statistics of $\phi$, determined by Eq.~(\ref{basic2}).

It is convenient to deal with the solution of equation~(\ref{basic1}) for Fourier transforms $\tilde\varpi(\bm k)$ and $\tilde \phi(\bm k)$ of the fields $\varpi$ and $\phi$. The solution is given by
\begin{equation}
\tilde\varpi(t,\bm k) =\int d\tau\, G(t-\tau,\bm k) \tilde \phi(\tau,\bm k),
\label{convol1}
\end{equation}
where $G(t, \bm k)$ is Green function, which satisfies the following differential equation
\begin{eqnarray}
\left(\frac{\partial}{\partial t}
+\alpha+\nu \bm k^2\right){G}(t,\bm k)
=\delta(t).
\label{gref1}
\end{eqnarray}
It can be easily solved to obtain
\begin{eqnarray}
G(t,\bm k)=\theta(t) \exp(-\alpha t-\nu \bm k^2 t),
\label{zerot1}
\end{eqnarray}
where $\theta(t)$ is the Heaviside step function reflecting causality.

Next, let us introduce the Fourier transform of the pair correlation function of vorticity
\begin{eqnarray}
\langle \varpi(t,\bm x) \varpi(0,\bm y)\rangle_0
=\int \frac{d^2k}{(2\pi)^2}
e^{i\bm k \cdot (\bm x- \bm y)}
{F}(t,\bm k),
\label{fouri2}
\end{eqnarray}
where the subscript $0$ means the linear approximation. Substituting expression~(\ref{convol1}) for $\varpi$ in terms of $\phi$ into Eq.~(\ref{fouri2}) and averaging the product $\varpi(t,\bm x) \varpi(0,\bm y)$ in accordance with Eq.~(\ref{basic2}), we obtain
\begin{eqnarray}
{F}(t, \bm k)=
2\epsilon\int\limits_{-\infty}^{\min(t,0)} d\tau\,
k^2 \tilde\varXi(\bm k)
{ G}(t-\tau,\bm k)
{ G}(-\tau,\bm k),
\label{fouri4}
\end{eqnarray}
where
\begin{equation}
\tilde\varXi(\bm k)=\int d^2 x\, \exp(-i\bm k \cdot \bm x) \varXi(\bm x)
\label{tildexi}
\end{equation}
is the Fourier transform of $\varXi(\bm r)$. Therefore, the simultaneous pair correlation function $F(\bm k) \equiv F(0, \bm k)$ equals
\begin{eqnarray}
F(\bm k)= \frac{\epsilon k^2}{\nu k^2 +\alpha}
\tilde\varXi(\bm k)
\label{isotrexp}
\end{eqnarray}
and
\begin{equation}
F(t,\bm k)=G(|t|,\bm k) F(\bm k).
\label{tdep}
\end{equation}
The last expression is a consequence of the short correlation in the time of our pumping.

The second moment of vorticity in the linear approximation is expressed as
\begin{eqnarray}
\langle \varpi^2 \rangle_0
=\int \frac{d^2 k}{(2\pi)^2} F(\bm k)
=\int \frac{d^2 k}{(2\pi)^2}
\frac{\epsilon k^2}{\alpha +\nu k^2}
\tilde\varXi(\bm k).
\label{secondm}
\end{eqnarray}
If $\alpha \sim \nu k_f^2$ then one derives from Eq. (\ref{secondm})
\begin{equation}
\langle \varpi^2 \rangle_0 \sim \epsilon/\nu
\sim \epsilon k_f^2 /\alpha.
\label{seczero}
\end{equation}
If $\alpha \ll \nu k_f^2$ then $\langle \varpi^2 \rangle_0 = \epsilon/\nu$ and in the opposite case $\alpha\gg \nu k_f^2$ one finds the estimate $\langle \varpi^2 \rangle_0 \sim \epsilon k_f^2 /\alpha$. All the values are proportional to the power (energy production rate) per unit mass $\epsilon$, with different factors.

The value of the second moment of the velocity can be found as
\begin{equation}
\langle v^2 \rangle_0
=\int \frac{d^2 k}{(2\pi)^2}
\frac{F(k)}{k^2}
=\epsilon \int \frac{d^2 k}{(2\pi)^2}
\frac{\tilde \varXi(k)}{\alpha+\nu k^2}.
\label{velocmom}
\end{equation}
In the limit $\alpha \gg \nu k_f^2$, the expression (\ref{velocmom}) gives $\langle v^2 \rangle_0=\epsilon/\alpha$. In the opposite limit $\alpha \ll \nu k_f^2$, the integral (\ref{velocmom}) is gained from the interval $\sqrt{\alpha/\nu}< k < k_f$. Then one obtains
\begin{equation}
\langle v^2 \rangle_0=
\frac{\epsilon \tilde \varXi(0)}{4\pi \nu}
\ln\frac{\nu k_f^2}{\alpha}.
\label{momlog}
\end{equation}
The expression assumes that $\tilde\Xi(k)$ can be approximated by $\tilde\Xi(0)$ inside the interval of the integration. Note that $\langle v^2 \rangle_0\sim k_f^{-2}\langle \varpi^2 \rangle_0$, up to the logarithmic factor at small $\alpha$.

One can also calculate the turbulent diffusion coefficient $D=\int dt\, \langle \bm v(t,\bm r) \cdot \bm v (0,\bm r)\rangle_0$,
\begin{eqnarray}
D=\frac{\epsilon}{2}\int \frac{d^2 k}{(2\pi)^2}
\frac{\tilde \varXi(k)}{(\alpha+\nu k^2)^2}.
\label{diffusion}
\end{eqnarray}
It determines an evolution of an impurity field if molecular diffusion is smaller than $D$. In the limit $\alpha \gg \nu k_f^2$, expression~(\ref{diffusion}) gives $D=\epsilon/(2\alpha^2)$. In the opposite limit $\alpha \ll \nu k_f^2$, we find
\begin{equation}
D=\frac{\epsilon \tilde \varXi(0)}{16\alpha \nu}
\sim \frac{\epsilon}{\alpha \nu k_f^2}.
\label{Dbroad}
\end{equation}
The expression also assumes that $\tilde\Xi(k)$ can be approximated by $\tilde\Xi(0)$ inside the interval of the integration. Note that $\alpha$ does not drop from the expression since it is determined by $k\sim \sqrt{\alpha/\nu}$.

Let us emphasize that results (\ref{momlog}) and (\ref{Dbroad}) were obtained for broadband pumping, and they need to be corrected for narrowband pumping, since in the latter case $\tilde\varXi(\bm 0)=0$. For the forcing with spectrum (\ref{narrow}), the integrals over $\bm k$ can be easily calculated, and we obtain
\begin{equation}
\langle v^2 \rangle_0
=\frac{\epsilon}{\alpha+\nu k_f^2},
\quad
D=\frac{\epsilon}{2(\alpha+\nu k_f^2)^2}.
\end{equation}
In the limit $\alpha \gtrsim \nu k_f^2$, the results coincide with previous estimates, the differences arise only in the limit of small bottom friction $\alpha \ll \nu k_f^2$.

In order to find corrections to the linear approximation, one has to take into account the nonlinear term $\bm v \cdot \nabla \varpi$ in the equation (\ref{basic}). Corrections to the correlation functions, found in the linear approximation, can be obtained as a series over the nonlinear term. The corresponding technique is presented in the next section. The corrections can be expected to be weak if one of the two dimensionless parameters is small
\begin{equation}
\beta_\nu=\frac{\epsilon}{\nu^{3} k_f^{4}}, \qquad
\beta_\alpha=\frac{\epsilon k_f^2}{\alpha^3}.
\label{smallp}
\end{equation}
The first parameter $\beta_\nu$ in Eq.~(\ref{smallp}) is a power of Reynolds number taken at the pumping scale. The second parameter $\beta_\alpha$ in Eq.~(\ref{smallp}) is related to bottom friction and characterizes its strength at the same pumping scale. Further we check the expectation, that the parameters (\ref{smallp}) control the perturbation series, and examine an interplay of viscosity and bottom friction.

\section{Functional Integrals}\label{sec:perturb}

The nonlinear interaction of the flow fluctuations can be consistently examined in the framework of Wyld diagrammatic technique \cite{Wyld61}. The diagrammatic technique can be derived from the representation of correlation functions as functional integrals over the observed variables and auxiliary fields \cite{MSR73}. The integration is performed like in the quantum field theory \cite{qufield}. A detailed description of the technique can be found in the review~\cite{HRS}.

In our case, in order to find correlation functions of the field $\varpi$, one should average over the solutions of equation~(\ref{basic}). However, it is convenient to integrate over arbitrary functions $\varpi(t,\bm r)$ taking into account the equation (\ref{basic}) by the corresponding functional $\delta$-function. Then $\delta$-function can be written as the functional integral over the auxiliary scalar field $\mu$ of the corresponding exponent. After averaging the exponent over the statistics of external forcing by using Eq.~(\ref{basic2}), we obtain the functional integral with the weight $\exp(-{\mathcal I})$, where ${\mathcal I}$ is the effective action. It is given by the sum of two terms
\begin{eqnarray}
{\mathcal I}={\mathcal I}_2+{\mathcal I}_{int},
\label{gener1}
\end{eqnarray}
where
\begin{eqnarray}
{\mathcal I}_2= \int dt\, d^2x\, \mu (\partial_t+\alpha -\nu \nabla^2)\varpi
\nonumber \\
+\epsilon \int dt\, d^2x\, d^2 r\, \nabla^2\varXi(\bm x-\bm r) \mu(t,\bm x) \mu(t,\bm r),
\label{gener2} \\
{\mathcal I}_{int}=
\int dt\, d^2x\, \mu \bm v \cdot \nabla \varpi .
\label{generint}
\end{eqnarray}
The velocity $\bm v$ in Eq. (\ref{generint}) is implied to be expressed via the vorticity $\varpi$, see Sec.~\ref{sec:general}.

Next, any correlation function of the fields $\varpi$ and $\mu$ can be written as the functional integral. For example, the pair correlation function of vorticity is given by
\begin{equation}
\langle \varpi(t, \bm x) \varpi (0, \bm y)\rangle
= \int D\varpi\, D\mu\, e^{-{\mathcal I}}
 \varpi(t, \bm x) \varpi (0, \bm y).
 \label{paircofu}
\end{equation}
We also introduce the Green function (or the response function) in the coordinate representation
\begin{equation}
\langle \varpi(t, \bm x) \mu (0, \bm y)\rangle
= \int D\varpi\, D\mu\, e^{-{\mathcal I}}
 \varpi(t, \bm x) \mu (0, \bm y),
 \label{greenofu}
\end{equation}
which determines the linear response of the system to an additional small external force. Indeed, let us assume that some additional term $\Delta \phi=\mathrm{curl}\, \Delta \bm f$ is added to the right-hand side of Eq.~(\ref{basic}). Then the corresponding term $\Delta{\mathcal I}$ should be also added to the effective action,
\begin{equation}
\Delta{\mathcal I}=-\int dt\, d^2x\, \mu \Delta\phi.
\label{addition}
\end{equation}
Without this term $\langle \varpi \rangle = \int D\varpi\, D\mu\, e^{-{\mathcal I}} \varpi = 0$, but as a response to the additional forcing some non-zero average $\Delta\langle \varpi \rangle$ may appear. It can be written as the functional integral
\begin{equation}
\Delta\langle \varpi \rangle= \int D\varpi\, D\mu\, e^{-{\mathcal I}-\delta{\mathcal I}} \varpi.
\end{equation}
Expanding the exponent in $\Delta{\mathcal I}$, we find the linear response
\begin{equation}
\Delta\langle \varpi(t,\bm r) \rangle=
\int d\tau\, d^2x\, \langle \varpi(t,\bm r) \mu(\tau,\bm x) \rangle
\Delta \phi(\tau,\bm x),
\label{response}
\end{equation}
which is just the definition of the Green function. Due to causality, the average $\langle \varpi(t,\bm r) \mu(\tau,\bm x) \rangle$ is equal to zero if $t<\tau$.

Analogously to Eqs.~(\ref{paircofu}) and (\ref{greenofu}), one can introduce the triple correlation function of vorticity
\begin{eqnarray}
\langle \varpi(t,\bm r)\varpi(\tau,\bm x)\varpi(s, \bm y) \rangle
\nonumber \\
=\int D\varpi\, D\mu\, e^{-{\mathcal I}}
\varpi(t,\bm r)\varpi(\tau,\bm x)\varpi(s, \bm y).
\label{ttriple}
\end{eqnarray}
The quantity is non-zero since there is no symmetry of the effective action $\mathcal I$, forbidding its non-zero value. However, one should be careful with moments of vorticity. It changes its sign under reflection. Reflection relative to the second coordinate axis implies the following transformation
\begin{eqnarray}
x_1\to -x_1, \quad x_2\to x_2, \quad
v_1\to -v_1,
\nonumber \\
v_2\to v_2, \quad
\varpi \to - \varpi, \quad \mu\to - \mu.
\label{reflection}
\end{eqnarray}
The effective action (\ref{gener1}) is invariant under the transformation (\ref{reflection}). Therefore all odd moments of vorticity, including the third moment, are zero. However, the triple correlation function of vorticity (\ref{ttriple}) is nonzero, since it changes its sign together with changing sign of the first components of the coordinates.

Note that the correlation functions of all orders containing only the auxiliary field $\mu$ are equal to zero. The reason is that the functional integral $\int D\varpi\, \exp(-{\mathcal I})$ gives the functional $\delta$-function of the field $\mu$ due to the structure of the effective action (\ref{gener1}). More detailed explanation based on discretization of functional integrals can be found in Ref.~\cite[Sec.~3]{HRS}.

To make a connection with the linear approximation discussed in Sec.~\ref{sec:linear}, one has to neglect the third-order term ${\mathcal I}_{int}$ (\ref{generint}) in the functional integrals (\ref{paircofu}) and (\ref{greenofu}). Then we arrive to the Gaussian functional integrals determining the ``bare'' correlation functions
\begin{eqnarray}
\langle \varpi(t,\bm x) \mu(0,\bm y)\rangle_0
=\int {\mathcal D}\varpi {\mathcal D}\mu\, e^{-{\mathcal I}_2}
\varpi(t,\bm x) \mu(0,\bm y),
\label{gener3} \\
\langle \varpi(t,\bm x) \varpi(0,\bm y)\rangle_0
=\int {\mathcal D}\varpi {\mathcal D}\mu\, e^{-{\mathcal I}_2}
\varpi(t,\bm x) \varpi(0,\bm y),
\label{genar3}
\end{eqnarray}
which can be found explicitly. The corresponding calculations lead to the conclusion that $\langle \varpi(t,\bm x) \varpi(0,\bm y)\rangle_0$ is determined by Eq.~(\ref{fouri2}). It is explained by the fact that the second-order approximation for the effective action is equivalent to the linear approximation for Eq.~(\ref{basic}). Next,
\begin{eqnarray}
\langle \varpi(t,\bm x) \mu(0,\bm y)\rangle_0
=\int \frac{d^2k}{(2\pi)^2}
e^{i\bm k \cdot (\bm x-\bm y)}
{G}(t,\bm k),
\label{fouri1}
\end{eqnarray}
where $G(t, \bm k)$ is given by Eq.~(\ref{zerot1}). The expression (\ref{fouri1}) becomes almost obvious if we compare the relations (\ref{convol1}) and (\ref{response}).

\section{Perturbation Theory}\label{sec:firstcorr}

Corrections to the bare values of the correlation functions are given by the perturbation series. They are obtained by the expansion of the factor $\exp({\mathcal I}_2+{\mathcal I}_{int})$ in the Taylor series over ${\mathcal I}_{int}$ in the corresponding functional integral. Each term of the expansion can be found analytically, using Wick theorem \cite{Wick50}: the average of a product of the fields $\varpi$ and $\mu$ is equal to the sum of products of the bare pair averages (\ref{gener3}) and (\ref{genar3}) organized by all possible pairings. As a result, we come to multiple integrals over times and wave vectors of some expressions that are products of (\ref{zerot1}) and (\ref{tdep}). One should also introduce factors, corresponding to gradients in $\nabla\varpi$ and to converting $\bm v \to \varpi$.

Let us demonstrate, how it is possible to obtain an analytical expression for corrections to the correlation functions and their correspondence to the diagrams. As an example, we consider the first correction to the correlation function $\langle \varpi \mu \rangle$, that is of the second-order in ${\mathcal I}_{int}$. Substituting the sum (\ref{gener1}) into Eq.~(\ref{greenofu}) and expanding up to the second-order in ${\mathcal I}_{int}$, one finds the correction
\begin{eqnarray}
\delta \langle \varpi(t,\bm x) \mu(0,\bm y) \rangle
=\frac{1}{2}\int D\varpi\, D\mu\,
\exp(-{\mathcal I}_2)
\nonumber \\
\varpi(t,\bm x)
{\mathcal I}_{int}^2 \mu(0,\bm y).
\end{eqnarray}
The integral is Gaussian, therefore its value can be found in accordance with the Wick theorem~\cite{Wick50}. Then, using expression~(\ref{generint}), we obtain
\begin{eqnarray}
\delta \langle \varpi(t,\bm x) \mu(0,\bm y) \rangle
=\int d\tau\, ds\, d^2 r\, d^2z\,
\nonumber \\
\left\{\langle \varpi(t,\bm x) \mu(\tau,\bm r)\rangle_0
\langle v_\beta(s,\bm z) \mu(0,\bm y)\rangle_0 \right.
\nonumber \\
\left[\langle v_\alpha(\tau,\bm r) \mu(s,\bm z)\rangle_0
\langle \partial_\alpha \varpi(\tau,\bm r) \partial_\beta \varpi(s,\bm z)\rangle_0 \right.
\nonumber \\ \left.
+\langle v_\alpha (\tau,\bm r) \partial_\beta \varpi(s,\bm z)\rangle_0
\langle \partial_\alpha \varpi(\tau,\bm r)  \mu(s,\bm z)\rangle_0 \right]
\nonumber \\
+\langle \varpi(t,\bm x) \mu(\tau,\bm r)\rangle_0
\langle \partial_\beta\varpi(s,\bm z) \mu(0,\bm y)\rangle_0
\nonumber \\
\left[\langle v_\alpha(\tau,\bm r) \mu(s,\bm z) \rangle_0
\langle \partial_\alpha \varpi(\tau,\bm r) v_\beta(s,\bm z)\rangle_0 \right.
\nonumber \\ \left. \left.
+ \langle v_\alpha(\tau,\bm r) v_\beta(s,\bm z)\rangle_0
\langle \partial_\alpha \varpi(\tau,\bm r)\mu(s,\bm z)  \rangle_0
\right] \right\}.
\end{eqnarray}
Next, passing to the Fourier representation and using the relations (\ref{varpi}) and (\ref{stre1}), we find
\begin{eqnarray}
\delta G(t,\bm k)=\int d\tau\, ds \, \frac{d^2 q}{(2\pi)^2}
V(-\bm q,\bm q+\bm k) V(\bm q, \bm k)
\nonumber \\
G(t-\tau,\bm k) F(\tau-s,\bm q)
G(\tau-s,\bm q+\bm k) G(s,\bm k),
\label{corrrr}
\end{eqnarray}
where $G$ and $F$ are determined by Eqs.~(\ref{zerot1}) and (\ref{tdep}), and
\begin{equation}
V(\bm k,\bm q)=\frac{1}{2}\left(\frac{1}{q^2}-\frac{1}{k^2}\right)
(q_2 k_1 -q_1k_2).
\label{turb1}
\end{equation}

\begin{figure}[t]
\includegraphics[width=0.9\linewidth]{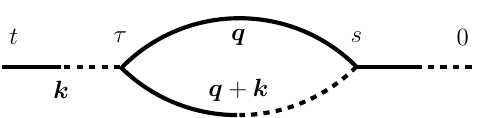}
\caption{One-loop diagram representing the first correction (\ref{corrrr}) to the Green function $G(t,\bm k)$.}
\label{fig:greencorr}
\end{figure}

\begin{figure}[t]
\includegraphics[width=0.45\linewidth]{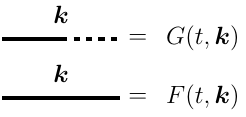}
\caption{Correspondence between the lines on diagrams and pair correlation functions.}
\label{fig:paircorrf}
\end{figure}

\begin{figure}[t]
\includegraphics[width=0.5\linewidth]{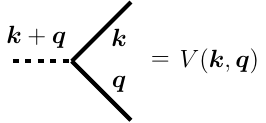}
\caption{Two solid segments and one dashed segment are attached to the vertex.}
\label{fig:vertex}
\end{figure}

Expression (\ref{corrrr}) corresponds to the diagram, which is shown in Fig.~\ref{fig:greencorr}. The lines on the diagrams correspond to the pair correlation functions, see Fig.~\ref{fig:paircorrf}. Each line consists of two segments, a solid segment corresponds to the field $\varpi$ and a dashed segment corresponds to the field $\mu$. The vertices on the diagrams are of the third order, as it is dictated by the structure of ${\mathcal I}_{int}$ (\ref{generint}). Two solid segments and one dashed segment are attached to each vertex, see Fig.~\ref{fig:vertex}. The analytical factors, corresponding to the vertices, are determined by Eq.~(\ref{turb1}).

\begin{figure}[t]
\includegraphics[width=0.9\linewidth]{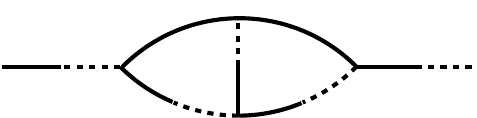}
\caption{Two-loop diagram representing a correction to the Green function.}
\label{fig:greencorr2}
\end{figure}

Next corrections to the correlation functions can be represented by Feynman diagrams as well. For example, one of the diagrams determining the second-order correction to Green function is shown in Fig.~\ref{fig:greencorr2}. Higher-order corrections correspond to more complicated diagrams. Note that the first-order corrections correspond to one-loop diagrams, the second-order corrections correspond to two-loop diagrams and so further.

The diagrams enable one to write easily the corresponding analytical expressions, using the rules depicted in Figs.~\ref{fig:paircorrf}, \ref{fig:vertex}. To construct the analytical expression, corresponding to a given diagram, one should fix the combinatorial factor, take the product of the factors, corresponding to the lines and the vertices, and then integrate the product over ``internal'' wave vectors and times. The integration should be performed at the condition of the wave vector conservation at each vertex: the sum of the wave vectors of three segments attached to a vertex has to be zero.

\subsection{Correction to the pair correlation function}
\label{sec:paircorr}

\begin{figure}
\includegraphics[width=0.9\linewidth]{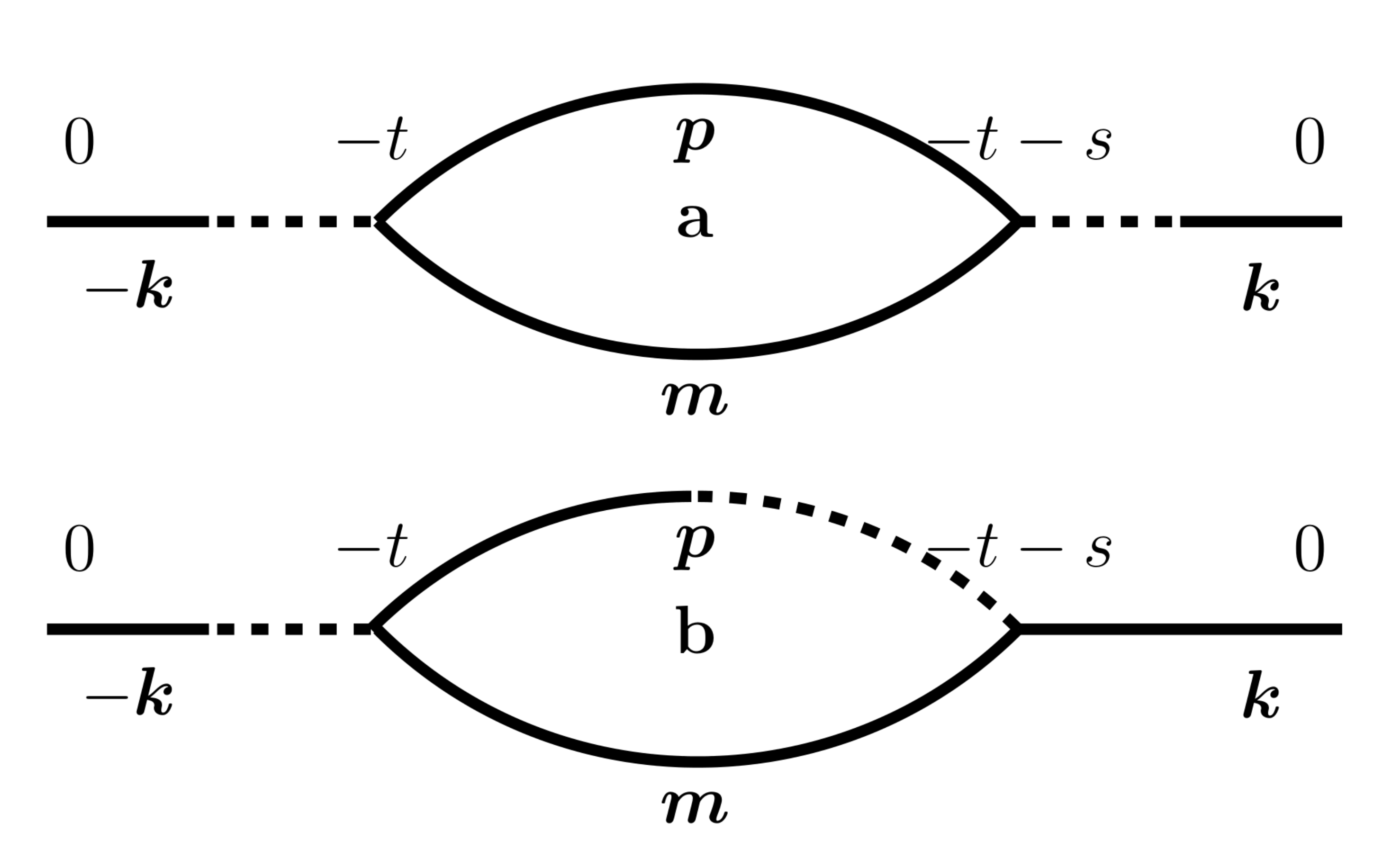}
\caption{First-order contributions to the simultaneous pair correlation function.}
\label{fig:paircorr}
\end{figure}

Here we examine the first correction $\delta F(\bm k)$ to the simultaneous pair correlation function $F(\bm k)$. The correction contains two contributions, determined by the diagrams depicted in Fig.~\ref{fig:paircorr}, labelled by the letters {\bf a} and {\bf b}. The diagrams correspond to the following analytical expressions
\begin{eqnarray}
F_{\mathrm a}(\bm k)
=\frac{1}{2}\int^{\infty}_0 dt \int_{-t}^\infty ds \int \frac{d^2 m}{(2\pi)^2}
\nonumber \\
G(t,\bm k) G(|s|,\bm p)
G(|s|,\bm m)  G(s+t, \bm k)
F(\bm m) F(\bm p)
\nonumber \\
\left(\frac{1}{m^2}-\frac{1}{p^2}\right)^2
(m_2k_1-m_1 k_2)^2,
\label{isotr6}
\end{eqnarray}
and
\begin{eqnarray}
F_{\mathrm b}(\bm k)
=-2\int_{-\infty}^0 dt \int_{0}^\infty ds \int \frac{d^2 m}{(2\pi)^2}
\nonumber  \\
G(t,\bm k) G(s,\bm p)
G(s,\bm m)  G(s+t, \bm k)
F(\bm m) F(\bm k)
\nonumber \\
\left(\frac{1}{m^2}-\frac{1}{p^2}\right)
\left(\frac{1}{\bm m^2}-\frac{1}{{\bm k}^2}\right)
(m_2k_1-m_1 k_2)^2,
\label{isotr7}
\end{eqnarray}
where $\bm p=-\bm k-\bm m$ and we substituted expression~(\ref{tdep}).

After substituting the expressions (\ref{zerot1}) and (\ref{isotrexp}) one can take the integrals over the times $t,s$ in Eqs.~(\ref{isotr6}) and (\ref{isotr7}) to obtain
\begin{eqnarray}
F_{\mathrm a}(\bm k)
=\frac{\epsilon^2}{2(\alpha+\nu k^2)} \int \frac{d^2 m}{(2\pi)^2}
\tilde \varXi(\bm m) \tilde \varXi (\bm p)
\nonumber \\
\frac{m^2 p^2}{(\alpha+\nu m^2)(\alpha+\nu p^2)(3\alpha+\nu k^2 +\nu p^2 +\nu m^2)}
\nonumber \\
\left(\frac{1}{m^2}-\frac{1}{p^2}\right)^2
(m_2k_1-m_1 k_2)^2,
\label{isotr8}
\end{eqnarray}
and
\begin{eqnarray}
F_{\mathrm b}(\bm k)
=-\frac{\epsilon^2}{\alpha+\nu k^2} \int \frac{d^2 m}{(2\pi)^2}
\tilde \varXi(\bm m) \tilde \varXi (\bm k)
\nonumber \\
\frac{m^2 k^2}{(\alpha+\nu m^2)(\alpha+\nu k^2)(3\alpha+\nu k^2 +\nu p^2 +\nu m^2)}
\nonumber \\
\left(\frac{1}{m^2}-\frac{1}{p^2}\right)
\left(\frac{1}{m^2}-\frac{1}{{k}^2}\right)
(m_2k_1-m_1 k_2)^2.
\label{isotr9}
\end{eqnarray}
The general expressions (\ref{isotr8}) and (\ref{isotr9}) enable one to analyze the correction $\delta F=F_{\mathrm a}+F_{\mathrm b}$ to the pair correlation function and to establish its behavior in different limiting cases.

Let us integrate over $\bm k$ the product $(\alpha+\nu k^2)\delta F(\bm k)$. One finds from Eq. (\ref{isotr8})
\begin{eqnarray}
\int \frac{d^2 k}{(2\pi)^2} (\alpha+\nu k^2) F_{\mathrm a}(\bm k)
=\epsilon^2 \int \frac{d^2 p\,d^2 m}{(2\pi)^4}
\tilde \varXi(\bm m) \tilde \varXi (\bm p)
\nonumber \\
\frac{p^2}{(\alpha+\nu m^2)(\alpha+\nu p^2)(3\alpha+\nu k^2 +\nu p^2 +\nu m^2)}
\nonumber \\
\left(\frac{1}{m^2}-\frac{1}{p^2}\right)
(m_2p_1-m_1 p_2)^2, \qquad
\label{faint}
\end{eqnarray}
where we passed to the integration over $\bm p$ and used symmetric properties of the integrand. The second integral is equal to
\begin{eqnarray}
\int \frac{d^2 k}{(2\pi)^2} (\alpha+\nu k^2)F_{\mathrm b}(\bm k)
=-\epsilon^2 \int \frac{d^2k\,d^2 m}{(2\pi)^4}
\tilde \varXi(\bm m) \tilde \varXi (\bm k)
\nonumber \\
\frac{1}{(\alpha+\nu m^2)(\alpha+\nu k^2)(3\alpha+\nu k^2 +\nu p^2 +\nu m^2)}
\nonumber \\
\left[k^2-\frac{k^2m^2}{p^2}\right]
\left(\frac{1}{m^2}-\frac{1}{{k}^2}\right)
(m_2k_1-m_1 k_2)^2. \quad
\end{eqnarray}
Here the first contribution, related to the term $k^2$ in the square brackets, is cancelled by the integral (\ref{faint}), and the second contribution, related to the term $k^2 m^2/p^2$, is zero since the integrand changes its sign under the permutation $\bm m \leftrightarrow \bm k$. We conclude that
\begin{equation}
\int \frac{d^2 k}{(2\pi)^2} (\alpha+\nu k^2)\delta F(\bm k)=0.
\label{enstrophy2}
\end{equation}
The relation is a direct consequence of the enstrophy balance (\ref{enstrophyb}), since the integral (\ref{enstrophy2}) is the correction of the order of $\epsilon^2$ to the right-hand side of relation (\ref{enstrophyb}), and the expression (\ref{enstrophy}) shows that such correction should be absent. Thus, our calculations are in agreement with the general relation (\ref{enstrophyb}).

Another check can be extracted from the energy balance (\ref{energyb}). It demonstrates that the correction of the order of $\epsilon^2$ to the right-hand side of relation~(\ref{energyb}) should be absent
\begin{equation}
\int \frac{d^2 k}{(2\pi)^2} \left(\frac{\alpha}{k^2}+\nu\right)\delta F(\bm k)=0.
\label{energyb2}
\end{equation}
Indeed,
\begin{eqnarray}
\int \frac{d^2 k}{(2\pi)^2} \left(\frac{\alpha}{k^2}+\nu\right)F_{\mathrm a}(\bm k)
=\epsilon^2 \int \frac{d^2 p\,d^2 m}{(2\pi)^4}
\tilde \varXi(\bm m) \tilde \varXi (\bm p)
\nonumber \\
\frac{(-m^2/k^2)}{(\alpha+\nu m^2)(\alpha+\nu p^2)(3\alpha+\nu k^2 +\nu p^2 +\nu m^2)}
\nonumber \\
\left(\frac{1}{m^2}-\frac{1}{p^2}\right)
(m_2p_1-m_1 p_2)^2, \qquad
\label{faint2}
\end{eqnarray}
and
\begin{eqnarray}
\int \frac{d^2 k}{(2\pi)^2}  \left(\frac{\alpha}{k^2}+\nu\right)F_{\mathrm b}(\bm k)
=-\epsilon^2 \int \frac{d^2k\,d^2 m}{(2\pi)^4}
\tilde \varXi(\bm m) \tilde \varXi (\bm k)
\nonumber \\
\frac{1}{(\alpha+\nu m^2)(\alpha+\nu k^2)(3\alpha+\nu k^2 +\nu p^2 +\nu m^2)}
\nonumber \\
\left[1-\frac{m^2}{p^2}\right]
\left(\frac{1}{m^2}-\frac{1}{{k}^2}\right)
(m_2k_1-m_1 k_2)^2. \quad
\end{eqnarray}
Here the first contribution, related to unity in the square brackets, is zero since the integrand changes its sign under the permutation $\bm m \leftrightarrow \bm k$, and the second contribution, related to the term $m^2/p^2$ in the square brackets, is cancelled by Eq.~(\ref{faint2}).

Next, let us consider the limiting case $\alpha\ll \nu k_f^2$. Putting $\alpha\to0$ in the expressions (\ref{isotr8}) and (\ref{isotr9}), we obtain
\begin{eqnarray}
F_{\mathrm a}(\bm k)
=\frac{\epsilon^2}{2\nu^4 k^2} \int \frac{d^2 m}{(2\pi)^2}
\frac{\tilde \varXi(\bm m) \tilde \varXi (\bm p)}{k^2 +p^2 +m^2}
\nonumber \\
\left(\frac{1}{m^2}-\frac{1}{p^2}\right)^2
(m_2k_1-m_1 k_2)^2,
\label{isotr10}
\end{eqnarray}
and
\begin{eqnarray}
F_{\mathrm b}(\bm k)
=-\frac{\epsilon^2}{\nu^4 k^2} \int \frac{d^2 m}{(2\pi)^2}
\frac{\tilde \varXi(\bm m) \tilde \varXi (\bm k)}{k^2 + p^2 + m^2}
\nonumber \\
\left(\frac{1}{m^2}-\frac{1}{p^2}\right)
\left(\frac{1}{\bm m^2}-\frac{1}{{\bm k}^2}\right)
(m_2k_1-m_1 k_2)^2.
\label{isotr11}
\end{eqnarray}
One can rewrite the sum $F_{\mathrm a}(\bm k)+F_{\mathrm b}(\bm k)$ as $A+B$, where
\begin{eqnarray}
A=\frac{\epsilon^2}{\nu^4 k^2} \int \frac{d^2 m}{(2\pi)^2}
\frac{\tilde \varXi(\bm m) \tilde \varXi (\bm p)}{k^2 + p^2 + m^2} (m_2k_1-m_1 k_2)^2
\nonumber \\
\left[-\frac{1}{2 m^4}+\frac{1}{2p^4} +\frac{1}{k^2 m^2} -\frac{1}{k^2 p^2} \right],
\label{zerocon}
\end{eqnarray}
and
\begin{eqnarray}
B=-\frac{\epsilon^2}{\nu^4 k^2} \int \frac{d^2 m}{(2\pi)^2}
\frac{\tilde \varXi(\bm m)[ \tilde \varXi (\bm k)-\tilde \varXi(\bm p)]}{k^2 + p^2 + m^2}
\nonumber \\
\left(\frac{1}{m^2}-\frac{1}{p^2}\right)
\left(\frac{1}{ m^2}-\frac{1}{{ k}^2}\right)
(m_2k_1-m_1 k_2)^2.
\label{isotr12}
\end{eqnarray}
The quantity $A$ is equal to zero, it can be checked performing the change of the integration variable $\bm m \to \bm p$ in Eq.~(\ref{zerocon}). Thus, the correction $\delta F$ is reduced to the value of $B$.

\begin{figure} \begin{center}
\includegraphics[width=0.9\linewidth]{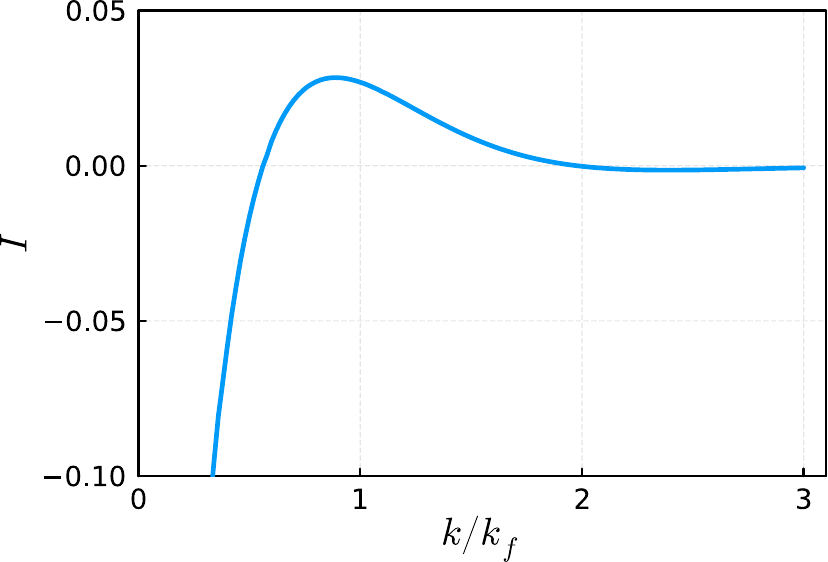} \end{center}
\caption{The integral (\ref{isotr14}) as a function of $k/k_f$.}
\label{fig:PCPI}
\end{figure}

There are no divergencies in integral (\ref{isotr12}). Therefore, we conclude that at $k\sim k_f$
\begin{equation}
\delta F(\bm k)=B
\sim \frac{\epsilon^2}{\nu^{4} k_f^{6}}
\sim \frac{\epsilon}{\nu^{3} k_f^{4}} F,
\label{smallpar1}
\end{equation}
where we used expression (\ref{isotrexp}). Thus, the small parameter of the perturbation series is the parameter $\beta_\nu$ (\ref{smallp}). Note that in the limit $\alpha\to 0$ the relations (\ref{enstrophy2}) and (\ref{energyb2}) are reduced to
\begin{equation}
\int d^2 k\, B(\bm k)=0, \quad
\int d^2 k\, k^2 B(\bm k)=0.
\label{zeroalpha}
\end{equation}
These relations can be checked directly, using the symmetry properties of integral~(\ref{isotr12}). In accordance with Eq.~(\ref{zeroalpha}), the function $B(k)$ should change its sign twice when varying $k$.

\begin{figure} \begin{center}
\includegraphics[width=0.9\linewidth]{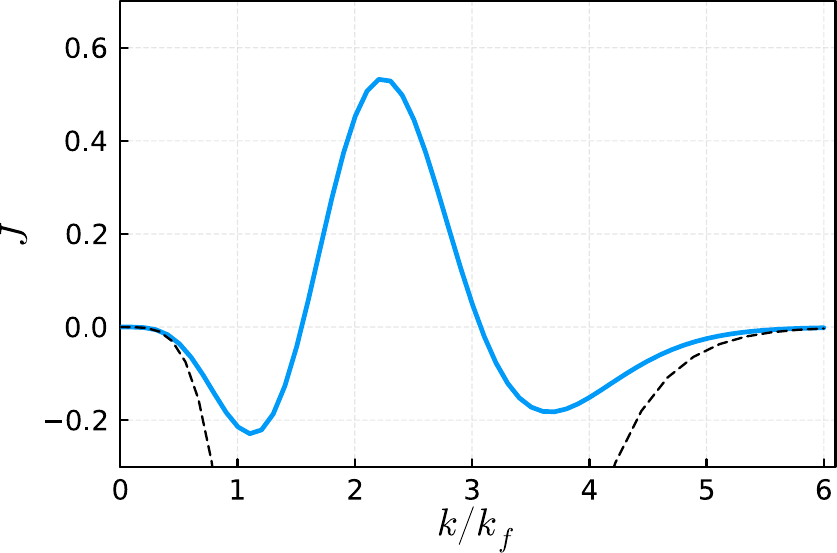} \end{center}
\caption{The integral (\ref{gralpha6}) as a function of $k/k_f$.}
\label{fig:PCLA}
\end{figure}

It is instructive to take a particular function $\varXi$, enabling to calculate integrals for the corrections up to a number. We will exploit function~(\ref{farce1}), which has Fourier transform~(\ref{gausspu}). Then expression~(\ref{isotr12}) is reduced to
\begin{eqnarray}
\delta F(\bm k)=B=
-\frac{\epsilon^2}{\nu^4 k_f^6} I,
\label{isotr13} \\
I=\int_0^\infty \frac{dm}{m}
\int_{-1}^1 dx \sqrt{1-x^2}
\nonumber  \\
\frac{k_f^2}{k^2 + m^2  -k m x}
\exp\left(-\frac{m^2+k^2}{2k_f^2}\right)
\nonumber \\
\left[1
-\exp\left(-\frac{m^2-2k m x}{2k_f^2}\right)\right]
\nonumber \\
\left(1-\frac{m^2}{k^2+m^2 -2 k m x}\right)
\left(1-\frac{m^2}{{ k}^2}\right).
\label{isotr14}
\end{eqnarray}
The integral converges and is, consequently, of the order of unity for $k\sim k_f$. Thus, the expression~(\ref{isotr13}) confirms the general evaluation~(\ref{smallpar1}).

The integral (\ref{isotr14}) (found numerically) is plotted in Fig.~\ref{fig:PCPI} as a function of $k/k_f$. In accordance with the general properties~(\ref{zeroalpha}), the function changes its sign twice when varying $k/k_f$. The integral has the following asymptotics
\begin{eqnarray}
I\approx -\frac{\pi}{8}\ln(k_f/k), \quad k\ll k_f,\\
I\approx -\frac{2^9 \pi k_f^8}{k^8}
\exp\left(-\frac{k^2}{4 k_f^2}\right), \quad k\gg k_f,
\label{asymptotic}
\end{eqnarray}
which can be found from Eq.~(\ref{isotr14}). The logarithmic dependence on $k$ is correct provided $k > \sqrt{\alpha/\nu}$; for smaller $k$, the bottom friction $\alpha$ cannot be neglected and the logarithm has to be replaced by $\ln(k_f \sqrt{\nu/\alpha})$. Note that the argument of the exponent is two times smaller than in Eq.~(\ref{gausspu}). It is a consequence of the fact that $\delta F$ is formed by nonlinear confluence of two fluctuations.

Next, we consider the opposite limiting case $\alpha\gg \nu k_f^2$. Then expressions (\ref{isotr8}) and (\ref{isotr9}) are reduced to
\begin{eqnarray}
F_{\mathrm a}(\bm k)
=\frac{\epsilon^2}{6\alpha^4} \int \frac{d^2 m}{(2\pi)^2}
\tilde \varXi(\bm m) \tilde \varXi (\bm p)
\nonumber \\
(p^2-m^2)
\left(\frac{1}{m^2}-\frac{1}{p^2}\right)
(m_2k_1-m_1 k_2)^2,
\label{gralpha3}
\end{eqnarray}
and
\begin{eqnarray}
F_{\mathrm b}(\bm k)
=-\frac{\epsilon^2}{3\alpha^4} \int \frac{d^2 m}{(2\pi)^2}
\tilde \varXi(\bm m) \tilde \varXi (\bm k)
\nonumber \\
\left(\frac{1}{m^2}-\frac{1}{p^2}\right)
(k^2-m^2)
(m_2k_1-m_1 k_2)^2.
\label{gralpha4}
\end{eqnarray}
Assuming $k\sim m \sim k_f$, we arrive at the estimate
\begin{equation}
F_{\mathrm a}\sim F_{\mathrm b} \sim \frac{\epsilon^2 k_f^2}{\alpha^4}
\sim \frac{\epsilon k_f^2}{\alpha^3} F,
\label{smallpar}
\end{equation}
where we used expression~(\ref{isotrexp}). Thus, the small parameter of the perturbation series in the limiting case $\alpha\gg \nu k_f^2$ is the parameter $\beta_\alpha$ (\ref{smallp}).

Note that in the limit $\nu\to 0$, relations (\ref{enstrophy2}) and (\ref{energyb2}) are reduced to
\begin{equation}
\int d^2 k\, \delta F(\bm k)=0, \quad
\int d^2 k\, k^{-2} \delta F(\bm k)=0.
\label{zeronu}
\end{equation}
These relations can be checked directly, using the symmetry properties of expressions (\ref{gralpha3}) and (\ref{gralpha4}). In accordance with Eq.~(\ref{zeronu}), the function $\delta F(k)$ should change its sign twice when varying $k$.

\begin{figure}[t]
\includegraphics[width=0.9\linewidth]{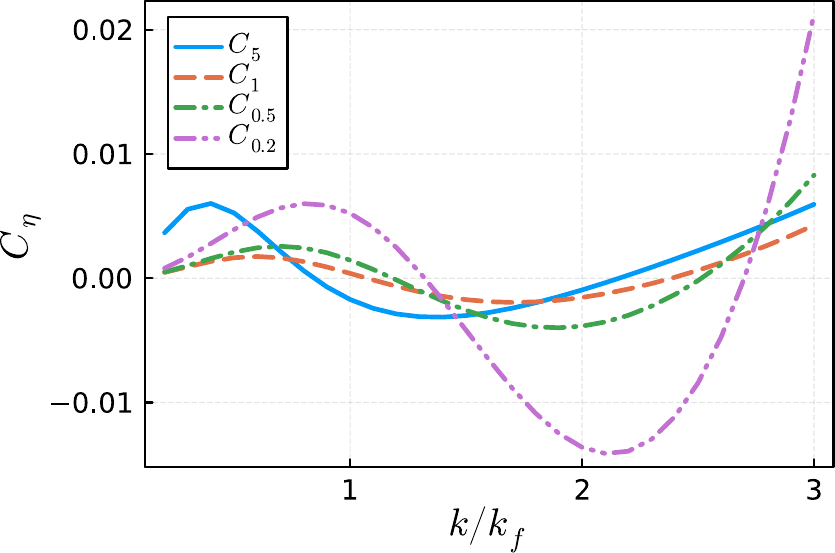}
\caption{Dependence of functions $C_{\eta}$ on parameter $k/k_f$ for different values of $\eta=\nu k_f^2/\alpha$.}
\label{fig:C_eta}
\end{figure}

Again, it is instructive to calculate the correction for a particular function $\varXi$. Using expression (\ref{gausspu}), one finds
\begin{equation}
F_{\mathrm a}+ F_{\mathrm b}
=-\frac{\epsilon^2 k_f^2}{3 \alpha^4} J,
\label{gralpha5}
\end{equation}
where
\begin{eqnarray}
J=\frac{k^2}{k_f^6}
\int_0^\infty dm\, m \int_{-1}^1 dx\, \sqrt{1-x^2}
\nonumber \\
\exp\left(-\frac{k^2+m^2}{2 k_f^2}\right)
\left\{2\frac{(k^2-m^2)(k^2-2kmx)}{k^2+m^2-2kmx}\right.
\nonumber \\ \left.
-\frac{(k^2-2kmx)^2}{k^2+m^2-2kmx}
\exp\left(-\frac{m^2}{2k_f^2}+ \frac{mkx}{k_f^2}\right) \right\}.
\label{gralpha6}
\end{eqnarray}
The integral converges, and thus $J\sim 1$ for $k\sim k_f$, confirming the general evaluation~(\ref{smallpar}).

The integral (\ref{gralpha6}) (found numerically) is plotted in Fig.~\ref{fig:PCLA} as a function of $k/k_f$. In accordance with the general properties (\ref{zeronu}), the function changes its sign twice when varying $k/k_f$. The asymptotics of the integral~(\ref{gralpha6})
\begin{eqnarray}
J\approx -\frac{\pi}{4}\frac{k^4}{k_f^4},  \quad k\ll k_f,\\
J\approx -8 \pi\exp\left(-\frac{k^2}{4 k_f^2}\right),
\quad k\gg k_f,
\end{eqnarray}
can be found directly from expression~(\ref{gralpha6}).

To handle an arbitrary ratio between parameters $\alpha$ and $\nu k_f^2$, based on expressions (\ref{smallpar1}) and (\ref{smallpar}), we introduce the following expression for the correction
\begin{equation}
\delta F(\bm k) = \dfrac{\beta_{\alpha} \beta_{\nu}}{\beta_{\alpha}
+\beta_{\nu}} F(\bm k) C_{\eta}(\bm k),
\label{defceta}
\end{equation}
where $\eta = \nu k_f^2/\alpha$. The function $C_{\eta}(\bm k)$ defined in this way can be found numerically using the expressions (\ref{isotr8}-\ref{isotr9}) and (\ref{isotrexp}). The results for forcing with the spectrum (\ref{gausspu}) are shown in Fig.~\ref{fig:C_eta}. It can be concluded that the values of $C_{\eta}$ are numerically small. In particular, as we will demonstrate below in Sec.~\ref{sec:directns} by direct numerical simulation, this will lead to the fact that perturbation theory predictions work well even for moderately large values of $\beta_{\nu}$ or $\beta_{\alpha}$ of the order of $10$.

Let us summarize results presented in this subsection. Eqs. (\ref{isotr6}) and (\ref{isotr7}) give general expressions for the first corrections to the pair correlation function of vorticity. When the pumping is shortly correlated in time, expressions (\ref{zerot1}), (\ref{isotrexp}) and (\ref{tdep}) can be used, which allows the corrections to be rewritten as (\ref{isotr8}) and (\ref{isotr9}). Next, Eq.~(\ref{isotr12}) provides the correction in the limit $\alpha \ll \nu k_f^2$. The final answer (\ref{isotr13}) can be obtained by specifying an explicit form (\ref{gausspu}) of the pumping correlation function. In the opposite limiting case $\alpha \gg \nu k_f^2$, the correction is given by Eqs. (\ref{gralpha3}) and (\ref{gralpha4}), and its explicit form (\ref{gralpha5}) can be found for the same forcing spectrum (\ref{gausspu}). For an arbitrary ratio between $\alpha$ and $\nu k_f^2$, we introduce expression (\ref{defceta}) for the correction to the pair correlation function of vorticity, which contains factor $C_{\eta}(\bm k)$. Numerical values of this factor can be obtained for the specific forcing model (\ref{gausspu}), see Fig.~\ref{fig:C_eta}.

Finally, let us add a few words about the case of a narrow pumping spectrum $\varXi(\bm k)$ in the Fourier space. In the extreme case (\ref{narrow}), both contributions $F_a$ and $F_b$ are equal to zero. When the spectrum has a finite width $\delta_f \ll k_f$, corrections $F_a$ and $F_b$ are no longer equal to zero, but contain an additional smallness compared to the estimates discussed earlier in this section, which is controlled by the ratio $\delta_f/k_f \ll 1$. It is interesting, that the contribution $F_b$ is localized near $k_f$, while the contribution $F_a$ determines the correction $\delta F(k)$ in the region $|k-k_f| \gg \delta_f$. These results will be illustrated by direct numerical simulation in Sec.~\ref{sec:directns}.

\subsection{Correction to the second moment of vorticity}
\label{sec:secondmom}

The second moment of vorticity $\langle \varpi^2 \rangle$ is the integral over the wave vector of its simultaneous pair correlation function. Therefore the first correction to the moment is
\begin{equation}
\delta\langle \varpi^2 \rangle
=\int \frac{d^2 k}{(2\pi)^2}
(F_{\mathrm a}+ F_{\mathrm b}),
\end{equation}
where $F_{\mathrm a}$ and $F_{\mathrm b}$ are corrections to the pair correlation function (\ref{isotr8}) and (\ref{isotr9}). If $\alpha\sim \nu k_f^2$ then we find the estimate
\begin{equation}
\delta\langle \varpi^2 \rangle
\sim \frac{\epsilon^2}{\nu^4 k_f^4}
\sim \frac{\epsilon^2 k_f^4}{\alpha^4},
\label{gensecmom}
\end{equation}
following from Eqs. (\ref{smallpar1}) and (\ref{smallpar}). However, if we are interested in the limiting cases $\nu k_f^2\gg\alpha$ or $\nu k_f^2\ll\alpha$, we cannot use expression~(\ref{isotr12}) or expressions (\ref{gralpha3}) and (\ref{gralpha4}) for the corrections to the pair correlation function since the integrals of the functions are equal to zero, see Eqs.~(\ref{zeroalpha}) and (\ref{zeronu}).

To calculate correction to the second moment, we will use general expressions (\ref{isotr8}) and (\ref{isotr9}). In the expression for $F_{\mathrm a}$, we pass from the integration over $\bm k$ to the integration over $\bm p$. In the expression for $F_{\mathrm b}$, we first symmetrize the integrand in $\bm k, \bm m$ and then permute $\bm k \leftrightarrow \bm p$. Collecting all terms, one finds
\begin{eqnarray}
\delta\langle \varpi^2 \rangle
=-\frac{\epsilon^2 \alpha}{2\nu^5}\int \frac{d^2 p\, d^2 m}{(2\pi)^4}
\nonumber \\
\frac{\tilde\varXi(\bm p)}{(p^2 +\alpha/\nu)^2}\
\frac{\tilde\varXi(\bm m)}{(m^2 +\alpha/\nu)^2}\
(p_1m_2-p_2m_1)^2
\nonumber \\
\frac{(m^2-p^2)^2}
{m^2 p^2(\bm m+\bm p)^2[\alpha/\nu+(\bm m+\bm p)^2]}
\nonumber \\
\frac{[(\bm m+\bm p)^2-m^2][(\bm m+\bm p)^2-p^2]}
{3\alpha/\nu+ p^2 +m^2 +(\bm m +\bm p)^2}.
\label{isotr2}
\end{eqnarray}
The expression is correct for an arbitrary function $\tilde\varXi$.

Now, let us consider the limiting case $\alpha \ll \nu k_f^2$. There is small factor $\alpha/(\nu k_f^2)$ in integral~(\ref{isotr2}), but one should be careful in estimating the correction. At small $\alpha$, there is an infrared contribution to integral~(\ref{isotr2}) coming from $p,m\sim \sqrt{\alpha/\nu}$. The contribution behaves $\propto \alpha^{-1}$ and, consequently, can compensate the small factor $\alpha$ in the integral. Since the infrared contribution to integral~(\ref{isotr2}) comes from $p,m\sim \sqrt{\alpha/\nu}\ll k_f$, we can substitute factors $\tilde\varXi$ by constants $\tilde\varXi(\bm 0)$. Then the infrared contribution to integral (\ref{isotr2}) appears to be proportional to
\begin{eqnarray}
\int d^2 m\, d^2 p\, d^2 q\,
\delta(\bm m +\bm p +\bm q)
\nonumber \\
\frac{(p_1m_2-p_2m_1)^2}{m^2 p^2 q^2}
\frac{1}{3\alpha/\nu+ p^2 +m^2 +q^2}
\nonumber \\
\left[\frac{a}{hc}-\frac{2}{c}+\frac{h}{ac}-\frac{a}{h^2}+\frac{1}{h}\right.
\nonumber \\ \left.
+\frac{1}{a}-\frac{h}{a^2}+\frac{c}{h^2}-\frac{2c}{ah}+\frac{c}{a^2}\right],
\end{eqnarray}
where $a=\alpha/\nu+p^2$, $h=\alpha/\nu+m^2$, $c=\alpha/\nu+q^2$. The first factor in the integrand is invariant under permutations of $\bm p, \bm m, \bm q$. Such permutation means the corresponding permutation of $a,h,c$. Performing the permutations we find that the second factor in the integrand (in the square brackets) is cancelled to zero. We conclude that the leading infrared contribution to integral (\ref{isotr2}) is zero.

However, there is a subleading infrared contribution to integral (\ref{isotr2}) determining the principal value of the correction to $\langle \varpi^2 \rangle$ due to the logarithmic divergence. To extract the principal value, one has to use the next term of the expansion $\tilde\varXi$ in the wave vectors:
\begin{equation}
\tilde\varXi(p)\tilde\varXi(m)
\to -\frac{2\pi^2 C}{k_f^6}(p^2+m^2).
\label{expansion}
\end{equation}
Here $C$ is the constant of the order of unity, it is equal to unity $C=1$ for the particular function (\ref{gausspu}). Substituting expression (\ref{expansion}) into integral (\ref{isotr2}), putting $\alpha\to0$ in the integrand and taking the integral over the angles, one finds
\begin{eqnarray}
\delta\langle \varpi^2 \rangle
=\frac{C\epsilon^2 \alpha}{4\nu^5k_f^6}\int_0^\infty \frac{d p\, d m}{ pm}
\frac{(p^2-m^2)^2}{(p^2+m^2)^2}
\nonumber \\
\frac{1}{b^4}\left\{12 +\frac{b^4 +4b^2-32}{\sqrt{4-b^2}}
-\frac{b^4+2b^2-4}{\sqrt{1-b^2}}\right\},
\end{eqnarray}
where $b=2pm(p^2+m^2)^{-1}$. Passing to the variables $b$, $S=p^2+m^2$, we find
\begin{eqnarray}
\delta\langle \varpi^2 \rangle
=\frac{C \epsilon^2 \alpha}{4\nu^5k_f^6}\int_0^\infty dS \int_0^1 db
\frac{\sqrt{1-b^2}}{S b}
\nonumber \\
\frac{1}{b^4}\left\{12 +\frac{b^4 +4b^2-32}{\sqrt{4-b^2}}
-\frac{b^4+2b^2-4}{\sqrt{1-b^2}}\right\}.
\label{usotr1}
\end{eqnarray}
The integral diverges logarithmically at small $S$ and $b$.

The logarithmic divergency is cut from below by the omitted terms with $\alpha$ in the integrand and from above by the wave vector $k_f$. Thus we arrive at the following regions of integration in Eq.~(\ref{usotr1})
\begin{equation}
\frac{\alpha}{\nu} <S<k_f^2, \quad
\left(\frac{\alpha}{\nu S}\right)^{1/2}<b<1.
\end{equation}
Keeping the main logarithmic term, one finds
\begin{eqnarray}
\delta\langle \varpi^2 \rangle
=-\frac{C \epsilon^2 \alpha}{2^{10}\nu^5k_f^6}
\left(\ln \frac{\nu k_f^2}{\alpha}\right)^2.
\label{usotr2}
\end{eqnarray}
Therefore we arrive at the estimate (up to the logarithmic factor)
\begin{equation}
\delta \langle \varpi^2 \rangle
\sim -\frac{\alpha \epsilon^2}{\nu^5 k_f^6}
\sim  -\frac{\alpha}{\nu k_f^2}
\beta_{\nu}
\langle \varpi^2 \rangle_0,
\label{isotr3}
\end{equation}
where $\langle \varpi^2 \rangle_0 \sim \epsilon/\nu$ is the zero order contribution to the second moment. We conclude that besides the factor $\beta_\nu$ (\ref{smallp}), controlling the perturbation series in the limit $\nu k_f^2 \gg \alpha$, the correction to the second moment contains the additional small factor $\alpha/(\nu k_f^2)$.

Next, we consider the opposite limiting case $\alpha\gg \nu k_f^2$. Then expression (\ref{isotr2}) can be rewritten as
\begin{eqnarray}
-\frac{\epsilon^2 \nu}{3\alpha^5}\int \frac{d p\, d m}{(2\pi)^2}pm (m^2-p^2)^2
{\tilde\varXi(p)} {\tilde\varXi( m)}
\nonumber \\
\int \frac{d\psi}{2\pi }\frac{\sin^2\psi}
{m^2+p^2 +2p m \cos\psi}
\nonumber \\
\left[p^2m^2+2(p^2+m^2)pm \cos\psi +4 p^2 m^2  \cos^2\psi\right],
\end{eqnarray}
where $\psi$ is the angle between $\bm p$ and $\bm m$, and we assumed $p\sim m\sim k_f$. Taking the integral over $\psi$, we find
\begin{eqnarray}
\delta \langle \varpi^2 \rangle
=-\frac{\epsilon^2 \nu}{12\alpha^5}\int \frac{d p\, d m}{(2\pi)^2}
{\tilde\varXi(p)} {\tilde\varXi( m)}
\nonumber \\
pm (m^2-p^2)^2
(m^2+p^2-|m^2-p^2|).
\label{gralpha1}
\end{eqnarray}
The integrand here is positive and has no singularities. Therefore we arrive at the estimate
\begin{eqnarray}
\delta\langle \varpi^2 \rangle
\sim - \frac{\epsilon^2 \nu k_f^6}{ \alpha^{5}}
\sim - \frac{\nu k_f^2}{\alpha} \beta_{\alpha} \langle \varpi^2 \rangle_0,
\label{gralpha2}
\end{eqnarray}
where we have taken into account the estimate $\langle \varpi^2 \rangle_0 \sim \epsilon k_f^2 /\alpha$. For the particular function (\ref{gausspu}) we find from Eq.~(\ref{gralpha1})
\begin{equation}
\delta\langle \varpi^2 \rangle=
- \frac{4\epsilon^2 \nu k_f^6}{3 \alpha^{5}}.
\end{equation}
The result confirms estimate~(\ref{gralpha2}). We conclude that besides the factor $\beta_\alpha$ (\ref{smallp}), controlling the perturbation series in the limit $\alpha\gg \nu k_f^2$, the correction to the second moment contains the additional small factor $\nu k_f^2/\alpha$.

Thus, we have established estimates for the correction to the second moment of vorticity $\delta\langle \varpi^2\rangle$ in both limiting cases, see Eqs.~(\ref{isotr3}) and (\ref{gralpha2}). They are anomalously small in accordance with the general properties following from the energy balance and the enstrophy balance. It is interesting that estimates (\ref{isotr3}) and (\ref{gralpha2}) contain both kinetic coefficients, $\nu$ and $\alpha$, whereas naively one may think that one of the kinetic coefficients would drop.

\section{Triple correlation function of vorticity}
\label{sec:triplecorr}

\begin{figure}
\includegraphics[width=0.8\linewidth]{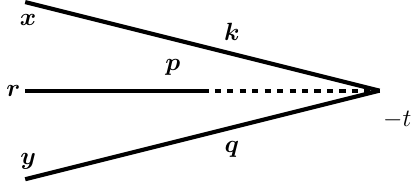}
\caption{One of the first order contributions to the triple correlation function of vorticity.}
\label{fig:thirdmom}
\end{figure}

In zero approximation, when one neglects the nonlinear interaction, the contribution to the triple correlation function of vorticity is equal to zero. Therefore the correlation function directly determines properties of the nonlinear interaction. Here we consider the first contribution to the triple correlation function of vorticity. It is determined by the first term in the expansion over the third-order term (\ref{generint}) in the effective action in the functional integral determining the triple correlation function of vorticity.

The first contribution to the triple correlation function of vorticity is determined by three similar diagrams, one of which is shown in Fig.~\ref{fig:thirdmom}. The corresponding analytical expression for the simultaneous triple correlation function is
\begin{eqnarray}
\langle \varpi(\bm x) \varpi(\bm y) \varpi(\bm r) \rangle =
\int_0^\infty dt \int \frac{d^2 k\, d^2q\, d^2 p}{(2\pi)^4}
\nonumber \\
\delta(\bm k+\bm q +\bm p)
\exp[i\bm k \cdot \bm x+i\bm q \cdot \bm y +i \bm p \cdot \bm r]
\nonumber \\
\left\{
\left(\frac{1}{q^2}-\frac{1}{k^2}\right)
(q_2 k_1 -q_1k_2)
F(t,\bm k) F(t,\bm q) G(t, \bm p)\right.
\nonumber \\
+\left(\frac{1}{k^2}-\frac{1}{p^2}\right)
(k_2 p_1 -k_1p_2)F(t,\bm k) F(t,\bm p)G(t,\bm q)
\nonumber \\ \left.
+\left(\frac{1}{p^2}-\frac{1}{q^2}\right)
(p_2 q_1 -p_1q_2)F(t,\bm p) F(t,\bm q)G(t,\bm k)
\right\}.
\label{third1}
\end{eqnarray}
Substituting here expression (\ref{tdep}) and taking the integral over time $t$, one finds
\begin{eqnarray}
\langle \varpi(\bm x) \varpi(\bm y) \varpi(\bm r) \rangle
=\int \frac{d^2 p\, d^2 q\, d^2 k}{(2\pi)^4}
\nonumber \\
\frac{\exp(i\bm k \cdot \bm x+i\bm q \cdot \bm y +i \bm p \cdot \bm r)}
{\nu(k^2+q^2+p^2)+3\alpha}
\delta(\bm k+\bm q+\bm p)
\nonumber \\
(q_2 k_1 -q_1k_2)\left\{
\left(\frac{1}{q^2}-\frac{1}{k^2}\right)
F(\bm q) F(\bm k)\right.
\nonumber \\
+\left(\frac{1}{k^2}-\frac{1}{p^2}\right)
F(\bm k) F(\bm p)
\nonumber \\ \left.
+\left(\frac{1}{p^2}-\frac{1}{q^2}\right)
F(\bm p) F(\bm q)\right\}.
\end{eqnarray}
The expression can be rewritten as
\begin{eqnarray}
\langle \varpi(\bm x) \varpi(\bm y) \varpi(\bm r) \rangle
=\varPhi(\bm x,\bm y,\bm r)
\nonumber \\
+\varPhi(\bm y,\bm r,\bm x)
+\varPhi(\bm r,\bm x,\bm y),
\label{third2}
\end{eqnarray}
where
\begin{eqnarray}
\varPhi(\bm x,\bm y,\bm 0)
=\int \frac{d^2 q\, d^2 k}{(2\pi)^4} \frac{\exp(i\bm k \cdot \bm x+i\bm q \cdot \bm y)}
{\nu(k^2+q^2+p^2)+3\alpha}
\nonumber \\
F(\bm q) F(\bm k)
(q_2 k_1 -q_1k_2)
\left(\frac{1}{q^2}-\frac{1}{k^2}\right).
\label{third3}
\end{eqnarray}
To restore $\varPhi(\bm x,\bm y,\bm r)$ from $\varPhi(\bm x,\bm y,\bm 0)$ one should substitute $\bm x \to \bm x - \bm r$, $\bm y \to \bm y-\bm r$ into $\varPhi(\bm x,\bm y,\bm 0)$.

Let us analyze the limiting case $\alpha \ll \nu k_f^2$. Then for the separations between points $\bm x,\bm y ,\bm r$ of the order of $k_f^{-1}$ we find from Eqs. (\ref{third2}) and (\ref{third3})
\begin{equation}
\langle \varpi \varpi \varpi \rangle
\sim \epsilon^2 \nu^{-3} k_f^{-2},
\label{thurd5}
\end{equation}
where we substituted $F\sim \epsilon/(\nu k_f^2)$. Thus, we find the estimate
\begin{equation}
\frac{\langle \varpi \varpi \varpi \rangle^2}
{(\langle \varpi^2 \rangle_0)^3}
\sim \frac{\epsilon}{\nu^3 k_f^4}.
\label{thurd4}
\end{equation}
The quantity in the right-hand side of Eq. (\ref{thurd4}) is the parameter $\beta_\nu$ (\ref{smallp}). We conclude that the perturbation parameter controls the third order correlation function in the limit $\alpha \ll \nu k_f^2$.

In the opposite limiting case $\alpha \gg \nu k_f^2$, we find for the separations $\sim k_f^{-1}$ from Eqs. (\ref{third2}) and (\ref{third3})
\begin{equation}
\langle \varpi \varpi \varpi \rangle
\sim \epsilon^2 k_f^4 \alpha^{-3},
\label{thurd1}
\end{equation}
where we substituted $F\sim \epsilon/\alpha$. Thus, we obtain
\begin{equation}
\frac{\langle \varpi \varpi \varpi \rangle^2}
{(\langle \varpi^2 \rangle_0)^3}
\sim \frac{\epsilon k_f^2}{\alpha^3}.
\label{thurd2}
\end{equation}
The quantity in the right-hand side of Eq. (\ref{thurd2}) is the parameter $\beta_\alpha$ (\ref{smallp}). We conclude that the perturbation parameter controls the third order correlation function in the limit $\alpha \gg \nu k_f^2$.

In the limit $\alpha \gg \nu k_f^2$, one can explicitly find $\varPhi(\bm x,\bm y,\bm 0)$ for the particular function (\ref{gausspu}), since in this case integral (\ref{third3}) is reduced to a Gaussian one. The result is
\begin{eqnarray}
\varPhi(\bm x,\bm y,\bm 0)=
\frac{\epsilon^2 k_f^4}{3 \alpha^3}
\left(\frac{\partial}{\partial x_1}\frac{\partial}{\partial y_2}
-\frac{\partial}{\partial y_1}\frac{\partial}{\partial x_2}\right)
\nonumber \\
\left[(x^2-y^2) \exp\left(-\frac{k_f^2 x^2}{2}
-\frac{k_f^2 y^2}{2}\right) \right].
\label{thurd3}
\end{eqnarray}
These expression is in agreement with the general estimate~(\ref{thurd1}).

We know that the third moment of vorticity is zero, see Section \ref{sec:perturb}. Therefore the triple correlation function should be small for the separations between the points $\bm x, \bm y, \bm r$ much smaller than $k_f^{-1}$. To analyze integral (\ref{third3}) in this case, we introduce the parametrization
\begin{eqnarray}
k_1= k \cos\varphi, \quad k_2=k \sin\varphi, \quad
q_1=q \cos(\varphi+\psi),
\nonumber \\
q_2=q\sin(\varphi+\psi), \quad
k_1q_2-k_2q_1=kq \sin\psi.
\nonumber
\end{eqnarray}
Taking the integral over $\varphi$, one obtains
\begin{eqnarray}
\varPhi(\bm x,\bm y,\bm 0)
=\int \frac{dq\, dk\, d\psi}{(2\pi)^3}
F(q) F(k)
\nonumber \\
\frac{(k^2-q^2)J_0(a) \sin\psi }
{2\nu(k^2+q^2+kq \cos\psi)+3\alpha},
\label{third4}
\end{eqnarray}
where
\begin{eqnarray}
a^2=k^2 x^2 +q^2 y^2 +2 k q \bm x \cdot \bm y \cos\psi
\nonumber \\
+2kq(x_1y_2-x_2y_1)\sin\psi.
\label{third5}
\end{eqnarray}
The quantity $a^2$ is positively defined.

Expanding $J_0$ in Eq.~(\ref{third4}) in $a^2$ and performing the integration we find a series over the separations. One can easily find that the terms of the expansion of zero, second, fourth and sixth orders in the separations produce zero contributions to $\langle \varpi(\bm x) \varpi(\bm y) \varpi(\bm r) \rangle$ in accordance with Eq.~(\ref{third2}). Thus, the first non-vanishing contribution to $\langle \varpi(\bm x) \varpi(\bm y) \varpi(\bm r) \rangle$ is of the eighth order in the separations. The property can be directly checked for the explicit expression~(\ref{thurd3}).

If $\alpha \ll \nu k_f^2$ then there is an interval of the separations between points
$k_f^{-1}\ll|\bm x - \bm y|, |\bm x - \bm r|, |\bm r - \bm y| \ll \sqrt{\nu/\alpha}$ where the triple correlation function of vorticity decays according to a universal power law. To find the law, one should note that for such distances the integral~(\ref{third3}) is gained at the wave vectors $k,q$ of the order of the inverse separations, $\sqrt{\alpha/\nu}\ll k,q\ll k_f$. Therefore, one can put $\alpha \to 0$ and expand the product $F(\bm k) F(\bm q)$ in $k^2$, $q^2$, since zero-order term of the expansion (that is a constant) leads to the function $\varPhi(\bm x,\bm y,\bm 0)$ producing zero contribution to the triple correlation function (\ref{third2}). Using equation (\ref{expansion}), we obtain
%Thus, we should take the first term of the expansion $F(\bm k) F(\bm q)$ in $k^2$, $q^2$, proportional to $k^2+q^2$. In accordance with Eq. (\ref{isotrexp}), in the limit $\alpha\to 0$ $F(k)\to (\epsilon/\nu)\tilde \varXi(k)$. The term of the expansion, needed for us, is determined by Eq. (\ref{expansion}). Substituting the expansion term into Eq. (\ref{third3}), calculating the function $\varPhi$ and substituting the result into Eq. (\ref{third2}) one finds
\begin{eqnarray}
\langle \varpi(\bm x) \varpi(\bm y) \varpi(\bm r) \rangle=
- C\frac{\epsilon^2}{(\nu k_f^2)^3}
\nonumber\\
\left[(x_1-r_1)(y_2-r_2)-(x_2-r_2)(y_1-r_1)\right]
\nonumber \\
\left\{\frac{4}{R^6}\left[\frac{(\bm x-\bm r)\cdot (2\bm y-\bm x-\bm r)}{(\bm x-\bm r)^2}-\right.\right.
\nonumber \\
\left.
\frac{(\bm y-\bm r)\cdot(2\bm x-\bm y-\bm r)}{(\bm y-\bm r)^2}-
\frac{(\bm x-\bm y)\cdot(2\bm r-\bm y-\bm x)}{(\bm y-\bm x)^2}\right]
\nonumber \\
 +\frac{1}{R^4}\left[\frac{(\bm x-\bm r)\cdot(2\bm y-\bm x-\bm r)}{(\bm x-\bm r)^4}-\right. \quad
 \label{ldi}\\
\left.\left.
\frac{(\bm y-\bm r)\cdot(2\bm x-\bm y-\bm r)}{(\bm y-\bm r)^4}-
\frac{(\bm x-\bm y)\cdot(2\bm r-\bm y-\bm x)}{(\bm y-\bm x)^4}\right]\right\},
\nonumber
\end{eqnarray}
where
\begin{equation}
R^2=(\bm x-\bm r)^2+(\bm y-\bm r)^2+(\bm x-\bm y)^2.
\nonumber
\end{equation}
The constant $C$ is equal to unity for the forcing with spectrum (\ref{gausspu}).

Expression (\ref{ldi}) is invariant under permutations of $\bm x,\bm y,\bm r$. Note that it is equal to zero when the points $\bm x,\bm y,\bm r$ are vertices of isosceles triangle. It is a consequence of the general symmetry: any odd correlation function of $\varpi$ should change its sign when reflected relative to an arbitrary axis. One can find a simple expression by putting $\bm r \to 0$ and assuming $x\ll y$. In this case, expression (\ref{ldi}) takes the form:
\begin{eqnarray}
\langle \varpi(\bm x) \varpi(\bm y) \varpi(\bm 0) \rangle=
\frac{\epsilon^2}{4(\nu k_f^2)^3}\frac{\sin (2\varphi)}{x^2 y^2},
\label{ldiapr}
\end{eqnarray}
where $\varphi$ is the angle between the vectors $\bm x$ and $\bm y$.

When separations are much larger than the length $\sqrt{\nu/\alpha}$, the triple correlation function $\langle \varpi(\bm x) \varpi(\bm y) \varpi(\bm r) \rangle$ decays exponentially. The argument of the exponent is determined by the same length $\sqrt{\nu/\alpha}$. Note that this behavior is universal, being independent of the pumping details.

\section{Direct Numerical Simulations}
\label{sec:directns}

In this section, we present the results of direct numerical simulations (DNS), which were performed to validate analytical expressions that underlie our analysis. The DNS results were obtained by integrating the Navier-Stokes equation in the vorticity formulation (\ref{basic}) using the GeophysicalFlows.jl pseudospectral code~\cite{Constantinou21}, fully dealiazed by the two-thirds rule. This code can be executed on the GPU, resulting in high computational performance. In our simulations, we used a periodic square domain of size $L=2\pi$.

\begin{figure}[t]
\includegraphics[width=0.9\linewidth]{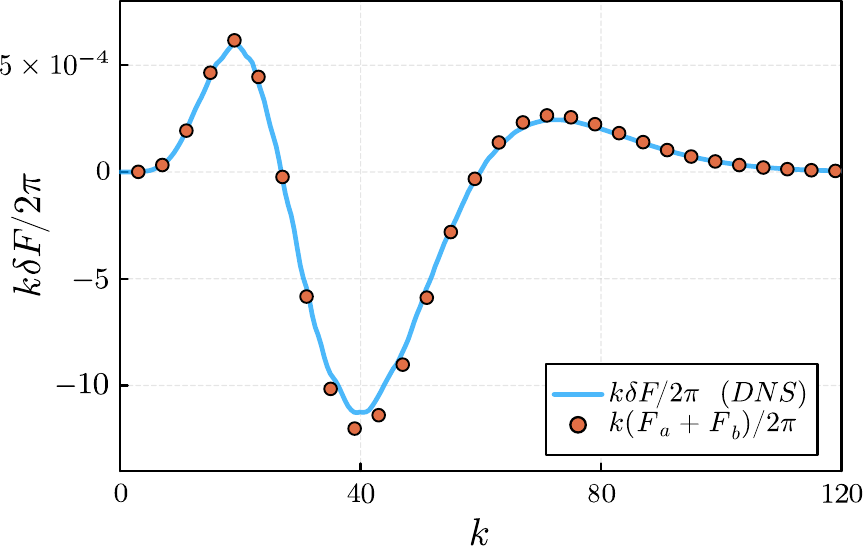}
\caption{Correction to the one-dimensional enstrophy spectrum. The DNS results (blue line) are consistent with the numerical integration of expressions (\ref{isotr8}) and (\ref{isotr9}) (markers). The values of parameters $\beta_{\nu}=\beta_{\alpha}=10$.}
\label{fig:DNS}
\end{figure}

We performed several runs to illustrate various features of obtained results. In the first run, our goal was to verify expressions (\ref{isotr8}) and (\ref{isotr9}), which describe corrections $F_a$ and $F_b$ to the pair correlation function of vorticity. We set the parameters $k_f=25$, $\nu = 1.6 \times 10^{-3}$ and $\alpha=1$, so that $\alpha=\nu k_f^2$. The random forcing was isotropic in space, shortly correlated in time and had a covariance spectrum (\ref{gausspu}). The grid resolution was $512^2$, which made it possible to collect the necessary statistics. Since $F_a,\, F_b \propto \epsilon^2$, we attempted to find the largest possible value of the parameter $\epsilon$ at which perturbation theory could still be applied. This will make it easier to distinguish the small correction $F_a+F_b$ from the background noise of the main contribution $F$, see expression (\ref{isotrexp}). We found that $\epsilon=1.6 \times 10^{-2}$ corresponding to the values of $\beta_{\nu}=\beta_{\alpha}=10$ leads to satisfactory results.

The initial state was a state of rest, and we first conducted simulations for some time to reach a statistical steady-state, observed by the saturation of the total kinetic energy. The integration time step was $\Delta t = 0.005$, which was also the correlation time of the exciting force. Once stationary, we saved the vorticity field and ran several simulations in parallel with different realizations of the random force, using the saved data as the initial condition. Then we stored data every $200$ integration steps and gathered a total of about $0.6 \times 10^6$ snapshots of the vorticity field.

Next, to compare DNS with analytical findings, we calculated the time-averaged one-dimensional enstrophy spectrum $k(F+\delta F)/(2\pi)$ and subtracted from it the contribution corresponding to the linear approximation (\ref{isotrexp}). The results are shown in Fig.~\ref{fig:DNS} and they agree with the answer obtained by numerical integration of expressions (\ref{isotr8}) and (\ref{isotr9}), which confirms their correctness. The slight noise noticeable in the DNS results is associated with a moderate volume of statistics.

\begin{figure}[t]
\includegraphics[width=0.9\linewidth]{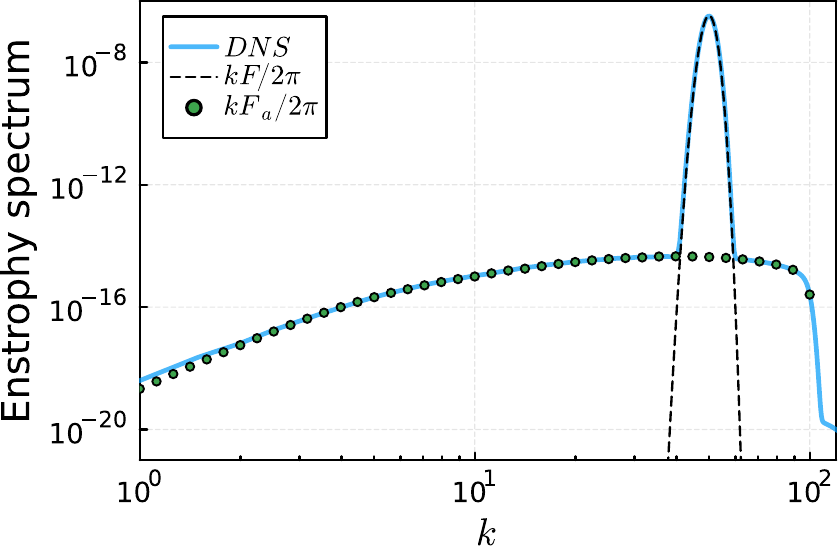}
\caption{Enstrophy spectrum under narrowband pumping (\ref{eq:pumping}) with $k_f=50$ and $\delta_f=1.5$. The solid line shows DNS results, the dashed line corresponds to the linear approximation (\ref{isotrexp}), and the markers are obtained by numerical integration of expression (\ref{isotr8}). The parameters $\beta_{\nu} = 10^{-3}$ and $\beta_{\alpha}=6.4 \times 10^{-2}$.}
\label{fig:DNS2}
\end{figure}

In the second run, our purpose was to considered the random forcing, which has a narrow spectrum in $k$-space. Pumping with such a spatial structure is commonly utilized in the numerical modeling of turbulent problems~\cite{Chertkov2007, Laurie2014, Frishman2018, Parfenyev2022} and, as mentioned in Section~\ref{sec:paircorr}, might result in some peculiarities. To illustrate the scenario, we performed DNS with
\begin{equation}\label{eq:pumping}
\tilde\varXi(\bm k) = \frac{\sqrt{2\pi}}{\delta_f k_f}
\exp \left[ - (k-k_f)^2/2 \delta_f^2 \right],
\end{equation}
where we set $k_f=50$ and $\delta_f=1.5$. The other parameters were $\alpha = 0.01$, $\nu=1.6 \times 10^{-5}$, $\epsilon=2.56 \times 10^{-11}$, which resulted in $\beta_{\nu} = 10^{-3}$ and $\beta_{\alpha}=6.4 \times 10^{-2}$, and the grid resolution was $1024^2$. The integration time step was $\Delta t = 0.01$ and after reaching a steady-state, we saved the vorticity field every $100$ integration steps, and collected $10^4$ snapshots.

The results for the time-averaged one-dimensional enstrophy spectrum $k(F+\delta F)/(2\pi)$ are presented in Fig.~\ref{fig:DNS2}. The linear approximation (\ref{isotrexp}) well describes the main peak near $k=k_f$, and the correction $F_a$ accounts for the wide plateau on which it is located. Note that in this case DNS does not require the collection of large statistics, since correction $F_a$ is separated from the main contribution in the Fourier space. The contribution $F_b$ is localized near $k_f$ and is not visible against the background of the main peak. These results confirm the qualitative analysis carried out in Subsection \ref{sec:paircorr}, and considering narrowband pumping allows us to double-check the correctness of the expression (\ref{isotr8}) with a few statistics in hand.

\begin{figure}[t]
\includegraphics[width=0.9\linewidth]{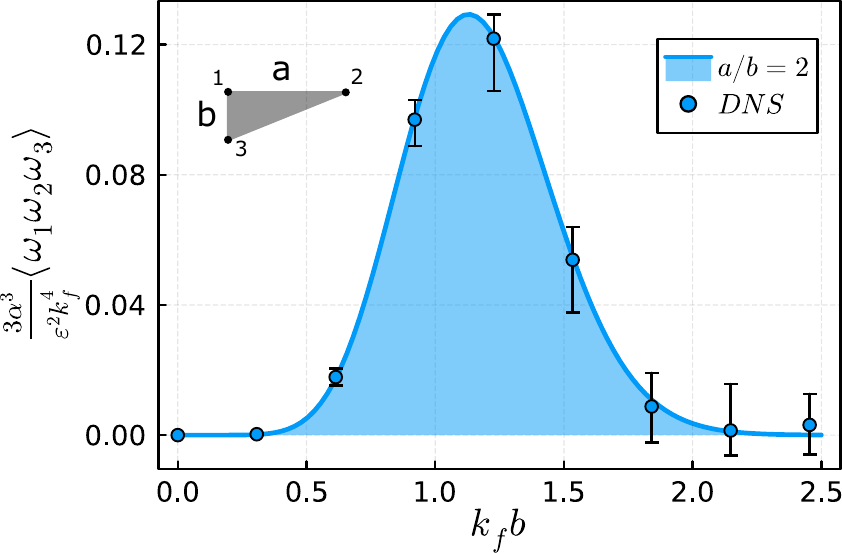}
\caption{Triple correlation function of vorticity: comparison of analytical expression (\ref{thurd3}) with DNS. The ratio $\nu k_f^2/\alpha = 10^{-3}$ and $\beta_{\alpha} = 10^{-4}$.}
\label{fig:DNS3}
\end{figure}

In the third run, we focused on triple correlation function of vorticity (\ref{third2}). Analytical calculations yielded a concise answer (\ref{thurd3}) in the limiting case $\nu k_f^2/\alpha \ll 1$ and when the pumping covariance spectrum had the form (\ref{gausspu}). We reproduced these conditions in DNS with parameters $\alpha = 1000$, $\nu=1.6 \times 10^{-3}$, $k_f=25$, $\epsilon=160$, which resulted in $\nu k_f^2/\alpha = 10^{-3}$ and $\beta_{\alpha} = 10^{-4}$. The grid resolution was $512^2$ and the integration time step was set to $\Delta t = 10^{-5}$. We started simulations from the state of rest, and once stationary, we saved the vorticity field and ran six simulations in parallel with different random realizations of forcing. We gathered a total of about $0.6 \times 10^6$ snapshots of the vorticity field, storing the data every $100$ integration steps.

Next, we calculated a triple correlation function $\langle \omega_1 \omega_2 \omega_3 \rangle$, where the values of vorticity were taken in the vertices of right triangle with legs $a$ and $b$. The results are presented in Fig.~\ref{fig:DNS3}. The solid lines show the analytical calculations using expression (\ref{thurd3}), and the markers represent the DNS results averaged over all collected data. The error bars indicate the maximum and minimum values obtained from six parallel simulations. Note that to increase statistics we used the symmetries of the problem: rotations of the triangle do not change the correlation function, and reflections relative to straight lines passing through the legs change its sign. Overall, it can be concluded that the analytical and numerical results are in good agreement with each other, which confirms the correctness of the developed perturbation method.

\section{Conclusion}
\label{sec:conclu}

We examined properties of the perturbation theory for two-dimensional fluid flow in the framework of the model where the pumping force, exciting the flow, is random in space and shortly correlated in time. The model admits advanced analytical calculations revealing qualitative properties of the perturbation theory. The peculiarity of two-dimensional turbulence is existence of two dissipation mechanisms: bottom friction and viscosity. There are two dimensionless parameters (\ref{smallp}) controlling the perturbation series for weak bottom friction and weak viscosity, correspondingly. We performed the calculations of the first correction to the pair correlation function of vorticity and the main contribution to the triple correlation function of vorticity, which enable one to evaluate the applicability region of the perturbation series.

There are some interesting features revealed by the calculations. Fourier transform of the correction to the pair correlation function changes its sign twice. It is the consequence of the integral relations (\ref{energyb2}) and (\ref{enstrophy2}), reflecting the energy balance and the enstrophy balance. The third moment of vorticity is zero in our model thanks to the symmetry reasoning, see Section \ref{sec:perturb}. At small separations between the points the triple correlation function of vorticity behaves as the eighth degree of the separations. If the bottom friction is weak then there is the universal region of distances where the triple correlation function scales as the power $-4$ of the distances, see expressions (\ref{ldi}) and (\ref{ldiapr}).

Surprisingly, the corrections to the second moment of vorticity do not obey the general properties of the perturbation series. In both limit cases, $\nu k_f^2\gg \alpha$ and $\nu k_f^2\ll \alpha$, the second moment is much less than the estimate, related to the corresponding small perturbation parameters (\ref{smallp}), see Sec. \ref{sec:secondmom}. Such behavior is explained by the energy and the enstrophy balances, leading to disappearing the main contributions to the correction to the second moment, see Sec. \ref{sec:secondmom}. Thus one should be careful with evaluating the parameter controlling the perturbation series via the second moment of vorticity.

To confirm our analytical results, we performed direct numerical simulations of two-dimensional turbulence excited by an external force shortly correlated in time. The simulations were performed in a square box with periodic boundary conditions. The results of the simulations are in excellent agreement with our analytical predictions both for corrections to the pair correlation function and for the triple correlation function. We tested also the pumping with a narrow spectrum in $\bm k$-space. The results are in agreement with the analytical calculations as well.

One can raise the question about the role of higher-order terms in the perturbation series: are they small provided that the first-order correction is small? Strictly speaking, we do not know the answer to the question. However, we can refer to the theory of quantum electrodynamics, which shows that the higher-order terms in the perturbation series are controlled by the same small parameter as the first-order correction. Say, the second-order contribution to the anomalous magnetic moment is near square of the first-order contribution~\cite{Akhiezer}. Hopefully, the situation is similar in our case.

One expects, that turbulence in a $2d$ film is excited where both parameters (\ref{smallp}) are large. The character of the state depends, generally, on the ratio $\alpha/(\nu k_f^2)$. However, large-scale fluctuations (with scales larger than the pumping length) are more sensitive to bottom friction whereas small-scale fluctuations (with scales smaller than the pumping length) are more sensitive to viscosity. Thus, the description of the turbulent state can be complicated, being dependent on the ratio $\nu k_f^2/\alpha$. The problem needs a special investigation.

The passive regime of the flow fluctuations where the perturbation series is applicable can be realized inside the coherent vortices appearing via the inverse cascade in two-dimensional turbulence \cite{KL15,kolokolov2016structure,kolokolov2016velocity,frishman17}. The reason is that the fluctuations are suppressed by the coherent vortex flow, which behaves locally as a shear flow. To establish the criterion of the applicability of the approach one should generalize the scheme developed in the present work to the case of the two-dimensional flow excited in the strong external shear flow. First steps in this direction were made in the papers \cite{KLT23a,KLT23b}.

\acknowledgments

I.V.K. acknowledges support from the Ministry of Science and Higher Education of the Russian Federation.
The work of V.V.L. and V.M.P. was supported by Grant 23-72-30006 of Russian Science Foundation. V.M.P. also acknowledges support from the Foundation for the Advancement of Theoretical Physics and Mathematics ``BASIS'', Project No. 22-1-3-24-1. Simulations were performed on the cluster of the Landau Institute for Theoretical Physics of RAS.

\end{document}